\documentclass[12pt]{article}
\usepackage{amsmath}
\usepackage{color}
\usepackage[unicode,bookmarks,bookmarksopen,bookmarksopenlevel=2,colorlinks,linkcolor=blue,citecolor=green]{hyperref}

\def\comment#1{}
\input{tcilatex}
\begin{document}

\title{Integrability of the Manakov--Santini hierarchy}
\author{Maxim V. Pavlov$^{1}$, Jen Hsu Chang$^{2}$, Yu Tung Chen$^{2}$ \\
$^{1}$Department of Mathematical Physics,\\
P.N. Lebedev Physical Institute of Russian Academy of Sciences,\\
Moscow, Leninskij Prospekt, 53\\
$^{2}$Department of Computer Science, National Defense University,\\
Taoyuan, Taiwan}
\date{}
\maketitle

\begin{abstract}
The first example of the so-called \textquotedblleft
coupled\textquotedblright\ integrable hydrodynamic chain is presented.
Infinitely many commuting flows are derived. Compatibility conditions of the
first two of them lead to the remarkable Manakov--Santini system.
Integrability of this four component three dimensional quasilinear system of
the first order as well as the coupled hydrodynamic chain is proved by the
method of hydrodynamic reductions. In comparision with a general case
considered by E.V. Ferapontov and K.R. Khusnutdinova, in this degenerate
case $N$ component hydrodynamic reductions are parameterized by $N+M$
arbitrary functions of a single variable, where $M$ is a number of branch
points of corresponding Riemann surface. These hydrodynamic reductions also
are written as symmetric hydrodynamic type systems. New classes of
particular solutions are found.
\end{abstract}

\tableofcontents

\bigskip

\textit{keywords}: hydrodynamic chains, Riemann invariants, symmetric
hydrodynamic type systems\textit{.}

\bigskip


\newpage

\section{Introduction}

This paper is dedicated to a new type of integrable systems, i.e. \textit{%
vector} hydrodynamic chains. The simplest example of \textit{scalar}
integrable hydrodynamic chains is the famous Benney hydrodynamic chain (see 
\cite{Benney})%
\begin{equation}
A_{t}^{k}=A_{x}^{k+1}+kA^{k-1}A_{x}^{0},\text{ \ \ }k=0,1,2,...  \label{raz}
\end{equation}%
This hydrodynamic chain can be written in the conservative form\footnote{%
as usual, a summation for \textit{negative} values of upper indices should
be ignored in all cases below.} (see, for instance, \cite{MaksGen})%
\begin{equation}
\partial _{t}H_{k}=\partial _{x}\left( H_{k+1}-\frac{1}{2}%
\sum_{m=0}^{k-1}H_{m}H_{k-1-m}\right) ,\text{ \ \ }k=0,1,2,...,  \label{dva}
\end{equation}%
where all conservation law densities $H_{k}$ are polynomial functions with
respect to moments $A^{n}$. For instance (see \cite{Benney}), $%
H_{0}=A^{0},H_{1}=A^{1},H_{2}=A^{2}+(A^{0})^{2},H_{3}=A^{3}+3A^{0}A^{1},...$

In this paper, we consider a simplest generalization of the Benney
hydrodynamic chain to a \textit{vector} case%
\begin{equation}
\left( 
\begin{array}{c}
q_{k} \\ 
f_{k}%
\end{array}%
\right) _{t}=\left( 
\begin{array}{c}
q_{k+1} \\ 
f_{k+1}%
\end{array}%
\right) _{x}-f_{0}\left( 
\begin{array}{c}
q_{k} \\ 
f_{k}%
\end{array}%
\right) _{x}-\sum_{m=0}^{k-1}q_{k-1-m}\left( 
\begin{array}{c}
q_{m} \\ 
f_{m}%
\end{array}%
\right) _{x},\text{ \ }k=0,1,2,...,  \label{j}
\end{equation}%
where the corresponding vector function contains two species of field
variables ($q_{k},f_{k}$) only. We believe that much more complicated
integrable \textit{vector} hydrodynamic chains%
\begin{equation}
\vec{A}_{t}^{k}=\hat{B}_{0}(\mathbf{A})\vec{A}_{x}^{k+1}+\hat{B}_{1}(\mathbf{%
A})\vec{A}_{x}^{k}+...+\hat{B}_{k+1}(\mathbf{A})\vec{A}_{x}^{0},\text{ \ }%
k=0,1,2,...  \label{vek}
\end{equation}%
can be discovered in coming future. In such a general construction, \textit{%
matrix} functions $\hat{B}_{k}(\mathbf{A})$ can depend on $\vec{A}^{0},\vec{A%
}^{1},...,\vec{A}^{k}$ for each index $k$ only, and vector $\vec{A}^{k}$ can
depend on an arbitrary number of components.

This vector hydrodynamic chain (\ref{j}) possesses two obvious reductions.
If $f_{k}$ vanish, then $q_{k}\rightarrow H_{k}$, and (\ref{j}) reduces to (%
\ref{dva}); if $g_{k}$ vanish, then (\ref{j}) reduces to the well-known
linearly degenerate hydrodynamic chain (see formula (37) in \cite{MaksEps})%
\begin{equation}
\left( f_{k}\right) _{t}=\left( f_{k+1}\right) _{x}-f_{0}\left( f_{k}\right)
_{x},\text{ \ }k=0,1,2,...  \label{cha}
\end{equation}%
Thus, (\ref{j}) is a natural generalization of two remarkable integrable
hydrodynamic chains on \textquotedblleft two species\textquotedblright\
case. This paper is particularly inspired by the first example in physical
literature such a Taranov's bi-chain (describing high frequency
electron-positron plasma waves, see detail in \cite{Taran})%
\begin{equation*}
A_{t}^{k}=A_{x}^{k+1}+kEA^{k-1},\text{ \ \ }B_{t}^{k}=B_{x}^{k+1}-kEB^{k-1},%
\text{ \ }k=0,1,2,...,
\end{equation*}%
where%
\begin{equation*}
dE=(A^{0}-B^{0})dx+(A^{1}-B^{1})dt.
\end{equation*}%
However, this bi-chain has no a \textquotedblleft
hydrodynamical\textquotedblright\ origin, i.e. this bi-chain does not
possess hydrodynamic reductions (see \cite{FK}). In a contrary with this
bi-chain, hydrodynamic bi-chain (\ref{j}) possesses infinitely many
hydrodynamic reductions. Moreover, hydrodynamic bi-chain (\ref{j}) is
connected with the three dimensional quasilinear Manakov--Santini system
determined by a commutativity condition of two vector fields (see \cite{MS}%
). In this paper, these vector fields are utilized for a derivation of
hydrodynamic bi-chain (\ref{j}). Method of symmetric hydrodynamic reductions
(see \cite{algebra}) is extended on this hydrodynamic bi-chain. A
construction of higher symmetries (see \cite{BCC}) is reformulated for a
description of hydrodynamic reductions, higher commuting flows and for an
application to the generalized (Tsarev's) hodograph method (see \cite{Tsar}%
). Thus, in this paper, we prove an integrability of hydrodynamic bi-chain (%
\ref{j}) by the method of hydrodynamic reductions adopted on two species of
moments $A^{k}$ and $B^{k}$. Moreover, extracting symmetric hydrodynamic
reductions (see \cite{algebra}), we present a list of infinitely many
explicit hydrodynamic reductions, i.e. one can construct infinitely many
particular solutions for the aforementioned Manakov--Santini system as well
as for hydrodynamic bi-chain (\ref{j}). First examples of three-component
hydrodynamic reductions of the Manakov--Santini system were found in \cite%
{ChangChen}. In comparison with the previous preliminary investigation (see 
\cite{ChangChen}), we been able to prove a consistency of all described
hydrodynamic reductions with hydrodynamic bi-chain (\ref{j}) as well as with
a whole Manakov--Santini hierarchy.

This paper is organized in the following way. In Section 2, a four component
three dimensional hydrodynamic type system (equivalent to the
Manakov--Santini system) is presented. Due to its dispersionless Lax
representation, a first example of integrable vector hydrodynamic chains is
derived. Simple reductions to well-known hydrodynamic chains are found. In
Section 3, a generating function of conservation laws and commuting flows is
described. In Section 4, the Manakov--Santini system is obtained from a new
hydrodynamic bi-chain. In Section 5, semi-Hamiltonian reductions of the
Manakov--Santini system are considered. It is proved that a corresponding
dispersion relation is degenerated. In Section 6, two species of Riemann
invariants are introduced. Full extended Gibbons--Tsarev system is derived.
In Section 7, all symmetric hydrodynamic reductions are described, and some
of them are explicitly found. In Section 8, the generalized hodograph method
is applied for a construction of infinitely many particular solutions for
the Manakov--Santini hierarchy. In Conclusion, a generalization of the
approach presented in this paper is discussed.

\section{Lax representation}

The function $\lambda (x,t,y;q)$ determined by the pair of linear equations%
\begin{equation}
\lambda _{t}=(q-a)\lambda _{x}-u_{x}\lambda _{q},\text{ \ \ \ \ }\lambda
_{y}=(q^{2}-aq-c)\lambda _{x}-(qu_{x}+u_{t})\lambda _{q}  \label{lin}
\end{equation}%
exists, because a compatibility condition $(\lambda _{t})_{y}=(\lambda
_{y})_{t}$ implies to the three dimensional hydrodynamic type system%
\begin{equation}
a_{t}=(c+u)_{x},\text{ \ \ }c_{t}=ca_{x}-ac_{x}-b_{x}+a_{y},\text{ \ \ }%
u_{t}=b_{x}-au_{x},\text{ \ \ }b_{t}=cu_{x}+u_{y},  \label{6}
\end{equation}%
which is equivalent to the well-known Manakov--Santini system (see \cite{MS})%
\begin{equation}
w_{xy}=w_{tt}+w_{x}w_{xt}+(u-w_{t})w_{xx},\text{ \ \ \ }%
u_{xy}=u_{tt}+u_{x}^{2}+w_{x}u_{xt}+(u-w_{t})u_{xx}.  \label{ms}
\end{equation}%
Indeed, the first two equations in (\ref{6}) reduce to the first equation in
(\ref{ms}) by the potential substitution $a=w_{x}$ and $c=w_{t}-u$, while
the second equation in (\ref{ms}) can be obtained from the compatibility
condition $(b_{t})_{x}=(b_{x})_{t}$ of other two equations in (\ref{6}).

An inverse transformation $\lambda (x,t,y;q)\rightarrow q(x,t,y;\lambda )$
leads to the pair of compatible quasilinear equations%
\begin{equation}
q_{t}=(q-a)q_{x}+u_{x},\text{ \ \ \ \ }q_{y}=(q^{2}-aq-c)q_{x}+qu_{x}+u_{t}.
\label{sem}
\end{equation}%
Let us introduce another compatible pair of \textit{auxiliary} quasilinear
equations (\textit{formally}, replacing $q_{x},q_{t},q_{y}$ and $u$ by $%
f_{x},f_{t},f_{y}$ and $a$, respectively)%
\begin{equation}
f_{t}=(q-a)f_{x}+a_{x},\text{ \ \ \ \ }f_{y}=(q^{2}-aq-c)f_{x}+qa_{x}+a_{t},
\label{osem}
\end{equation}%
where $f(x,t,y;q)$ depends on the parameter $\lambda $ implicitly via a
dependence $q(x,t,y;\lambda )$. The compatibility conditions $%
(q_{t})_{y}=(q_{y})_{t}$ and $(f_{t})_{y}=(f_{y})_{t}$ imply to (\ref{6}) in
both cases.

The reduction $a=0,c=-u$ reduces (\ref{6}) to the famous
Khokhlov--Zabolotskaya equation (also well known as a dispersionless limit
of the Kadomtsev--Petviashvili equation)%
\begin{equation}
u_{t}=b_{x},\text{ \ \ }b_{t}=u_{y}-uu_{x}  \label{dikp}
\end{equation}%
determined by compatibility condition (\ref{sem}) under the reduction $f=0$;
the reduction $u=0,b=0$ reduces (\ref{6}) to the remarkable equation
(recognized as an anti-self-duality reduction of the famous Yang-Mills
equation, see \cite{Dunaj})%
\begin{equation}
a_{t}=c_{x},\text{ \ \ }c_{t}=ca_{x}-ac_{x}+a_{y},  \label{uni}
\end{equation}%
determined by compatibility condition (\ref{osem}) under the reduction $q=0$.

Let us substitute an expansion ($\lambda \rightarrow \infty ,q\rightarrow
\infty $, according to a general theory of a relationship between dKP
equation (\ref{dikp}) and Benney hydrodynamic chain (\ref{dva}); see, for
instance, \cite{MaksGen})%
\begin{equation}
q=\lambda -\frac{q_{0}}{\lambda }-\frac{q_{1}}{\lambda ^{2}}-\frac{q_{2}}{%
\lambda ^{3}}-...  \label{inv}
\end{equation}%
in the first equation of (\ref{sem}). Then the \textit{first} part of
hydrodynamic bi-chain (\ref{j}) is derived, where $u=q_{0}$ and $b=q_{1}$. A
substitution of an expansion ($\lambda \rightarrow \infty ,f\rightarrow
\infty $, according to a general theory of a relationship between
quasilinear system (\ref{uni}) and hydrodynamic chain (\ref{cha}); see
formula (39) in \cite{MaksEps})%
\begin{equation}
f=-\frac{f_{0}}{\lambda }-\frac{f_{1}}{\lambda ^{2}}-\frac{f_{2}}{\lambda
^{3}}-...  \label{f}
\end{equation}%
in the first equation of (\ref{osem}) implies to the \textit{second} part of
hydrodynamic bi-chain (\ref{j}), where $a=f_{0}$ and $%
c=f_{1}-q_{0}-f_{0}^{2}/2$. Thus, \textit{whole} hydrodynamic bi-chain (\ref%
{j}) \textbf{cannot} be derived just from the first equation of (\ref{sem})
as well as from the first linear equation of (\ref{lin}) by an inverse
transformation ($\lambda \rightarrow \infty ,q\rightarrow \infty $, see (\ref%
{inv}))%
\begin{equation}
\lambda =q+\frac{A^{0}}{q}+\frac{A^{1}}{q^{2}}+\frac{A^{2}}{q^{3}}...,
\label{apart}
\end{equation}%
which is useful in a derivation of Benney hydrodynamic chain (\ref{raz})
under a corresponding reduction $f=0$ to the dKP case (see, for instance, 
\cite{MaksGen}). In such a case, another formal expansion ($\lambda
\rightarrow \infty ,M\rightarrow \infty $, the so-called Orlov function; see 
\cite{BCC})%
\begin{equation*}
M=\sum_{n=1}^{\infty }nt^{n-1}\lambda ^{n-1}+\sum_{n=1}^{\infty
}v_{n}\lambda ^{-n}
\end{equation*}%
for the first equation\footnote{%
It means that the function $M(\lambda ,\mathbf{t},\mathbf{v})$ satisfies (%
\ref{lin}) as well as the function $\lambda (q,\mathbf{A})$.} of (\ref{lin})
can be utilized in a derivation of the second part (see \cite{ChangChen}) of
hydrodynamic bi-chain (\ref{j}), where $p_{n}=(v_{n})_{x}$ are nothing else
but conservation law densities (see below). Here $t^{0}=x,t^{1}=t,t^{2}=y$.
All other \textit{higher} \textquotedblleft times\textquotedblright\ $t^{k}$
are associated with higher commuting hydrodynamic bi-chains (see below).
Instead this Orlov function $M$, in this paper, we use an alternative
approach (see (\ref{sem}) and (\ref{osem})) to derive a generating function
of conservation laws ($p\equiv M_{x}$, where $\lambda $ is a free parameter;
see detail in \cite{ChangChen}) as well as a generating function of
commuting flows (see below).

A substitution (\ref{apart}) to (\ref{f}) implies to the expansion ($%
f\rightarrow \infty ,q\rightarrow \infty $)%
\begin{equation}
f=-\frac{B^{0}}{q}-\frac{B^{1}}{q^{2}}-\frac{B^{2}}{q^{3}}-...  \label{non}
\end{equation}

Let us rewrite (\ref{osem}) splitting dynamics with respect to $x,t,y$ and $%
q $. It means the dynamics (\ref{sem}) is excluded from (\ref{osem}), and $q$
becomes an extra independent variable as well as $x,t,y$. Corresponding
equations reduce to the form (cf. (\ref{lin}))%
\begin{equation}
f_{t}=(q-a)f_{x}-u_{x}f_{q}+a_{x},\text{ \ \ }%
f_{y}=(q^{2}-aq-c)f_{x}-(qu_{x}+u_{t})f_{q}+qa_{x}+a_{t}.  \label{g}
\end{equation}

A substitution (\ref{apart}) to the first equation in (\ref{lin}) and (\ref%
{non}) to the first equation in (\ref{g}) leads to a generalization of
Benney hydrodynamic chain (\ref{raz}) to the vector case (cf. (\ref{j}))%
\begin{equation}
\left( 
\begin{array}{c}
A^{k} \\ 
B^{k}%
\end{array}%
\right) _{t}=\left( 
\begin{array}{c}
A^{k+1} \\ 
B^{k+1}%
\end{array}%
\right) _{x}-B^{0}\left( 
\begin{array}{c}
A^{k} \\ 
B^{k}%
\end{array}%
\right) _{x}+k\left( 
\begin{array}{c}
A^{k-1} \\ 
B^{k-1}%
\end{array}%
\right) A_{x}^{0},\text{ \ }k=0,1,2,...,  \label{bi}
\end{equation}%
where $a=B^{0}$ and $u=A^{0}$.

This hydrodynamic bi-chain has three obvious reductions. If $A^{k}=0$, then (%
\ref{bi}) reduces to (\ref{cha}); if $B^{k}=0$, then (\ref{bi}) reduces to (%
\ref{raz}); if $B^{k}=A^{k}$, then (\ref{bi}) reduces to the
\textquotedblleft deformed\textquotedblright\ Benney hydrodynamic chain%
\begin{equation*}
A_{t}^{k}=A_{x}^{k+1}-A^{0}A_{x}^{k}+kA^{k-1}A_{x}^{0},\text{ \ }k=0,1,2,...,
\end{equation*}%
derived in \cite{Haan}, whose integrability was investigated in \cite%
{MaksHam}. This hydrodynamic chain is connected (see \cite{Inter}) with the
so-called \textquotedblleft interpolating system\textquotedblright\ (see 
\cite{Dun}).

Moreover, Manakov--Santini system (\ref{ms}) possesses a natural reduction $%
u=w_{t}$ (see \cite{ChangChen}) to the continuum limit of the discrete KP
hierarchy (see, for instance, \cite{LeiYu})%
\begin{equation}
\Omega _{xy}=\Omega _{tt}+\frac{1}{2}\Omega _{xt}^{2},  \label{ek}
\end{equation}%
where $w=\Omega _{t}$. In such a case (\ref{6}) reduces (i.e. $c=0$) to%
\begin{equation*}
a_{t}=u_{x},\text{ \ \ }a_{y}=b_{x},\text{ \ \ }b_{t}=u_{y},\text{ \ \ }%
u_{t}=b_{x}-au_{x}.
\end{equation*}%
A comparison of the first conservation law $a_{t}=u_{x}$ (i.e. $%
B_{t}^{0}=A_{x}^{0}$, see the notation in (\ref{bi})) with the first
conservation law of (\ref{bi})%
\begin{equation*}
B_{t}^{0}=\left( B^{1}-\frac{1}{2}(B^{0})^{2}\right) _{x}
\end{equation*}%
implies to the \textit{first} constraint%
\begin{equation*}
A^{0}=B^{1}-\frac{1}{2}(B^{0})^{2}.
\end{equation*}%
It means, that hydrodynamic bi-chain (\ref{bi}) reduces to the hydrodynamic
chain%
\begin{equation}
B_{t}^{k}=B_{x}^{k+1}-B^{0}B_{x}^{k}+kB^{k-1}\left( B^{1}-\frac{1}{2}%
(B^{0})^{2}\right) _{x},  \label{be}
\end{equation}%
which is a particular case of a more general class of integrable
hydrodynamic chains ($\alpha ,\beta ,\gamma $ are arbitrary constants)%
\begin{equation*}
B_{t}^{k}=B_{x}^{k+1}-B^{0}B_{x}^{k}+(\alpha k+\beta )B^{k}B_{x}^{0}+\gamma
kB^{k-1}\left( B^{1}+\frac{\beta -\alpha -\gamma }{2}(B^{0})^{2}\right) _{x}
\end{equation*}%
investigated in \cite{KuperNorm} and then in \cite{MaksHam}. A computation
of all other constraints $A^{k}(B^{0},B^{1},...,B^{k+1})$ is a more
complicated task. Nevertheless, the second constraint can be found due to
comparison of the fourth equation $u_{t}=b_{x}-au_{x}$ (see (\ref{ek})) with
the first equation $A_{t}^{0}=A_{x}^{1}-B^{0}A_{x}^{0}$ from the $A$-part of
(\ref{bi}). It means that $b=A^{1}$ and a substitution of the first
constraint $A^{0}=B^{1}-(B^{0})^{2}/2$ leads to the conservation law $%
B_{t}^{1}=A_{x}^{1}$. Taking into account the second equation from (\ref{be}%
), the \textit{second} constraint $A^{1}=B^{2}-(B^{0})^{3}/3$ can be easily
obtained.

In aforementioned case (\ref{ek}), the first equation in (\ref{sem}) reduces
to%
\begin{equation*}
p_{t}=\partial _{x}(e^{p}+B^{0}),
\end{equation*}%
where $p=\ln (q-B^{0})$. This is a generating function of conservation laws
for the remarkable hydrodynamic chain (see, for instance, \cite{Stron} and 
\cite{LeiYu})%
\begin{equation*}
C_{t}^{k}=C_{x}^{k+1}+kC^{k}C_{x}^{0},\text{ \ \ }k=0,1,2,...,
\end{equation*}%
where $C^{0}=B^{0}$ and all other moments $C^{k}$ are connected with $B^{n}$
by inverse triangular point transformations (see detail in \cite{MaksHam}).
Since this hydrodynamic chain belongs to the Boyer--Finley (a continuum
limit of 2D Toda Lattice) hierarchy (see \cite{Kodama}), a corresponding
algebraic mapping is given by the expansion ($\lambda \rightarrow \infty
,q\rightarrow \infty ,p\rightarrow \infty $)%
\begin{equation*}
\lambda =e^{p}+C^{0}+e^{-p}C^{1}+e^{-2p}C^{2}+...=q+\frac{C^{1}}{q-C^{0}}+%
\frac{C^{2}}{(q-C^{0})^{2}}+\frac{C^{3}}{(q-C^{0})^{3}}+...,
\end{equation*}%
which must coincide with two other expansions ($\lambda \rightarrow \infty
,q\rightarrow \infty ;$ see (\ref{apart}))%
\begin{equation}
\lambda =q+\frac{A^{0}}{q}+\frac{A^{1}}{q^{2}}+\frac{A^{2}}{q^{3}}%
...=(q-B^{0})\exp \left( \frac{B^{0}}{q}+\frac{B^{1}}{q^{2}}+\frac{B^{2}}{%
q^{3}}...\right) .  \label{cons}
\end{equation}%
Thus, we see that hydrodynamic chain (\ref{be}) is embedded in hydrodynamic
bi-chain (\ref{bi}) by very complicated reductions $%
A^{k}(B^{0},B^{1},...,B^{k+1})$, while (\ref{ms}) reduces to (\ref{ek}) by
the simple constraint $c=0$ (see (\ref{6})). Moreover, in this case, both
linear systems (\ref{sem}) and (\ref{g}) coincide due to aforementioned link
(see (\ref{non}) and (\ref{cons}))%
\begin{equation*}
\lambda =(q-B^{0})e^{-f}.
\end{equation*}

The general problem of all possible compatible reductions $%
A^{k}(B^{0},B^{1},...,B^{k+n})$ should be considered in a separate
publication (for each fixed non-negative integer $n$).

\section{Generating functions}

Let us rewrite the first equation of (\ref{osem}) in the form%
\begin{equation*}
s_{t}=(q-a)s_{x}+sa_{x},
\end{equation*}%
where (see (\ref{f}))%
\begin{equation}
s=\exp f=1-\frac{s_{0}}{\lambda }-\frac{s_{1}}{\lambda ^{2}}-\frac{s_{2}}{%
\lambda ^{3}}-...  \label{exp}
\end{equation}%
Then hydrodynamic bi-chain (\ref{j}) reduces to the form%
\begin{equation*}
(s_{k})_{t}=(s_{k+1})_{x}-s_{0}(s_{k})_{x}-%
\sum_{m=0}^{k-1}q_{k-1-m}(s_{m})_{x}+s_{k}(s_{0})_{x},\text{ \ }%
(q_{k})_{t}=(q_{k+1})_{x}-s_{0}(q_{k})_{x}-%
\sum_{m=0}^{k-1}q_{k-1-m}(q_{m})_{x},
\end{equation*}%
where $s_{0}\equiv f_{0}$ (all other higher unknown functions $s_{k}$ are
polynomial functions of $f_{0},q_{0},f_{1}$, $q_{1},...,f_{k},q_{k}$ for
each index $k$). If $q_{k}$ vanish, this hydrodynamic bi-chain reduces
precisely to the form derived in \cite{Alonso}.

Let us introduce the so-called \textit{vertex} operator ($\zeta \rightarrow
\infty $)%
\begin{equation}
\partial _{\tau (\zeta )}=-\frac{1}{\zeta }\partial _{t^{0}}-\frac{1}{\zeta
^{2}}\partial _{t^{1}}-\frac{1}{\zeta ^{3}}\partial _{t^{2}}-...
\label{vertex}
\end{equation}

\textbf{Theorem}: \textit{Higher commuting hydrodynamic bi-chains are
determined by the pair of \textbf{vertex} equations}%
\begin{equation}
\partial _{\tau (\zeta )}s(\lambda )=\frac{s(\zeta )\partial _{x}s(\lambda
)-s(\lambda )\partial _{x}s(\zeta )}{q(\lambda )-q(\zeta )},\text{ \ \ \ }%
\partial _{\tau (\zeta )}q(\lambda )=s(\zeta )\partial _{x}\ln (q(\lambda
)-q(\zeta ))  \label{e}
\end{equation}%
\textit{by a substitution} (\ref{inv}), (\ref{exp}),%
\begin{equation*}
q(\zeta )=\zeta -\frac{q_{0}}{\zeta }-\frac{q_{1}}{\zeta ^{2}}-\frac{q_{2}}{%
\zeta ^{3}}-...,\text{ \ \ \ \ }s(\zeta )=1-\frac{s_{0}}{\zeta }-\frac{s_{1}%
}{\zeta ^{2}}-\frac{s_{2}}{\zeta ^{3}}-...
\end{equation*}%
\textit{and} (\ref{vertex}).

\textbf{Proof}: A direct substitution (\ref{inv}), (\ref{exp}) and (\ref%
{vertex}) in (\ref{e}) leads to a family of infinitely many hydrodynamic
bi-chains. Their commutativity follows from the compatibility conditions $%
\partial _{\tau (\zeta )}(\partial _{t}s(\lambda ))=\partial _{t}(\partial
_{\tau (\zeta )}s(\lambda )),\partial _{\tau (\zeta )}(\partial
_{y}s(\lambda ))=\partial _{y}(\partial _{\tau (\zeta )}s(\lambda
)),\partial _{\tau (\zeta )}(\partial _{t}q(\lambda ))=\partial
_{t}(\partial _{\tau (\zeta )}q(\lambda )),\partial _{\tau (\zeta
)}(\partial _{y}q(\lambda ))=\partial _{y}(\partial _{\tau (\zeta
)}q(\lambda ))$ leading to the extra \textit{vertex} dynamics (see (\ref{6}))%
\begin{eqnarray}
\partial _{\tau (\zeta )}a &=&\partial _{x}s(\zeta ),\text{ \ \ \ \ \ }%
\partial _{\tau (\zeta )}c=(q(\zeta )-a)\partial _{x}s(\zeta )-s(\zeta
)\partial _{x}(q(\zeta )-a),  \notag \\
&&  \label{ken} \\
\partial _{\tau (\zeta )}u &=&s(\zeta )\partial _{x}q(\zeta ),\text{ \ \ \ \
\ \ }\partial _{\tau (\zeta )}b=s(\zeta )\partial _{x}\left( \frac{%
q^{2}(\zeta )}{2}+u\right) ,  \notag
\end{eqnarray}%
which are compatible (i.e. $\partial _{\tau (\zeta )}(\partial
_{t}a)=\partial _{t}(\partial _{\tau (\zeta )}a),\partial _{\tau (\zeta
)}(\partial _{t}b)=\partial _{t}(\partial _{\tau (\zeta )}b),\partial _{\tau
(\zeta )}(\partial _{t}c)=\partial _{t}(\partial _{\tau (\zeta )}c),\partial
_{\tau (\zeta )}(\partial _{t}u)=\partial _{t}(\partial _{\tau (\zeta )}u)$)
with (\ref{6}). Thus, the theorem is proved.

\textbf{Remark}: This generating function of commuting flows (\ref{ken}) can
be obtained by a substitution (\ref{inv}), (\ref{exp}) in (\ref{e}) only.

Taking into account the first equation in (\ref{ken}), expansions (\ref{exp}%
) and (\ref{vertex}), the \textit{sole} function $S(t^{0},t^{1},t^{2},...)$
can be introduced such that%
\begin{equation*}
s(\zeta )=\partial _{\tau (\zeta )}S+1,\text{ \ \ \ }s_{k}=\frac{\partial S}{%
\partial t^{k}}.
\end{equation*}%
Then the first equation in (\ref{e}) implies to\footnote{%
here and below, the sign \textquotedblleft prime\textquotedblright\ means a
derivative with respect to a corresponding parameter.} 
\begin{equation*}
q^{\prime }(\lambda )=\frac{s(\lambda )\partial _{x}s^{\prime }(\lambda
)-s^{\prime }(\lambda )\partial _{x}s(\lambda )}{\partial _{\tau (\lambda
)}s(\lambda )}
\end{equation*}%
in the limit $\zeta \rightarrow \lambda $. Thus, all functions $q_{k}$ can
be expressed via first and second derivatives of the sole function $S$ with
respect to corresponding \textquotedblleft time\textquotedblright\ variables 
$t^{n}$. A substitution of these dependencies to the second equation in (\ref%
{e}) leads to an infinite set of three dimensional quasilinear equations of
the \textit{third} order. From the other hand, three copies of the first
equation in (\ref{e}) imply to a \textit{sole vertex} equation%
\begin{equation*}
\frac{\lbrack \partial _{\tau (\zeta )}S+1]\partial _{x}\partial _{\tau
(\lambda )}S-[\partial _{\tau (\lambda )}S+1]\partial _{x}\partial _{\tau
(\zeta )}S}{\partial _{\tau (\zeta )}\partial _{\tau (\lambda )}S}+\frac{%
[\partial _{\tau (\eta )}S+1]\partial _{x}\partial _{\tau (\zeta
)}S-[\partial _{\tau (\zeta )}S+1]\partial _{x}\partial _{\tau (\eta )}S}{%
\partial _{\tau (\eta )}\partial _{\tau (\zeta )}S}
\end{equation*}%
\begin{equation*}
+\frac{[\partial _{\tau (\lambda )}S+1]\partial _{x}\partial _{\tau (\eta
)}S-[\partial _{\tau (\eta )}S+1]\partial _{x}\partial _{\tau (\lambda )}S}{%
\partial _{\tau (\lambda )}\partial _{\tau (\eta )}S}=0.
\end{equation*}%
An expansion of this generating function with respect to three parameters $%
\lambda ,\zeta ,\eta $ at infinity leads to an infinite set of three
dimensional quasilinear equations of the \textit{second} order.

\textbf{Remark}: The second equation in (\ref{e}) implies to%
\begin{equation*}
s(\lambda )=\frac{\partial _{\tau (\lambda )}q(\lambda )}{\partial _{x}\ln
q^{\prime }(\lambda )}.
\end{equation*}%
in the limit $\zeta \rightarrow \lambda $. Thus, all functions $s_{k}$ can
be expressed via functions $q_{0},q_{1},...,q_{k}$ and their first
derivatives with respect to corresponding \textquotedblleft
time\textquotedblright\ variables $t^{n}$. A substitution of these
dependencies to the first equation in (\ref{e}) leads to an infinite set of
three dimensional quasilinear equations of the \textit{second} order.%
\begin{equation*}
\lbrack q(\lambda )-q(\zeta )]\partial _{\tau (\zeta )}\frac{\partial _{\tau
(\lambda )}q(\lambda )}{\partial _{x}\ln q^{\prime }(\lambda )}=\frac{%
\partial _{\tau (\zeta )}q(\zeta )}{\partial _{x}\ln q^{\prime }(\zeta )}%
\partial _{x}\frac{\partial _{\tau (\lambda )}q(\lambda )}{\partial _{x}\ln
q^{\prime }(\lambda )}-\frac{\partial _{\tau (\lambda )}q(\lambda )}{%
\partial _{x}\ln q^{\prime }(\lambda )}\partial _{x}\frac{\partial _{\tau
(\zeta )}q(\zeta )}{\partial _{x}\ln q^{\prime }(\zeta )}.
\end{equation*}%
Generating functions (\ref{e}) have an obvious two reductions. The choice $%
s(\lambda )=1$ means reduction to the generating function associated with
the dKP hierarchy (\ref{dikp}) and with the Benney hydrodynamic chain (\ref%
{dva}); the choice $q(\lambda )=\lambda $ means reduction to the generating
function associated with hydrodynamic chain (\ref{cha}), quasilinear system (%
\ref{uni}) and the so-called \textquotedblleft universal\textquotedblright\
hierarchy (see detail in \cite{Alonso} and \cite{MaksEps}).

\textbf{Theorem}: \textit{Hydrodynamic bi-chain} (\ref{j}) \textit{possesses
a generating function of conservation laws}%
\begin{equation}
\partial _{t}p=\partial _{x}[(q-a)p],  \label{geni}
\end{equation}%
\textit{where the generating function of conservation law densities is given
by}%
\begin{equation}
p(\lambda )=\frac{q^{\prime }(\lambda )}{s(\lambda )}.  \label{genu}
\end{equation}

\textbf{Proof}: Indeed, taking into account the first equation in (\ref{sem}%
) and the first equation in (\ref{osem}), a direct substitution (\ref{genu})
in (\ref{geni}) implies to an identity.

\textbf{Remark}: A substitution (\ref{inv}) and (\ref{exp}) in (\ref{genu})
leads to an expansion%
\begin{equation*}
p=1+\frac{p_{0}}{\lambda }+\frac{p_{1}}{\lambda ^{2}}+\frac{p_{2}}{\lambda
^{3}}+...,
\end{equation*}%
where all coefficients $p_{k}$ can be expressed polynomially via field
variables $q_{0},f_{0},q_{1},f_{1},...,q_{k},f_{k}$. Corresponding
conservation laws can be written in the form%
\begin{equation*}
p_{k,t}=\partial _{x}\Bigg(p_{k+1}-p_{0}p_{k}-q_{k}-%
\sum_{n=0}^{k-1}p_{k-n-1}q_{n}\Bigg),\text{ \ \ }k=0,1,2,...
\end{equation*}%
Of course, all conservation laws can be written via moments $A^{k},B^{k}$.
For instance, first three of them are%
\begin{equation*}
\partial _{t}B^{0}=\partial _{x}\left( B^{1}-\frac{(B^{0})^{2}}{2}\right) ,%
\text{ \ \ \ }\partial _{t}\left( B^{1}+\frac{(B^{0})^{2}}{2}+A^{0}\right)
=\partial _{x}\left( B^{2}-\frac{(B^{0})^{3}}{3}+A^{1}\right) ,
\end{equation*}%
\begin{eqnarray*}
&&\partial _{t}\left( B^{2}+B^{0}B^{1}+\frac{(B^{0})^{3}}{6}%
+2A^{1}+2A^{0}B^{0}\right) \\
&=&\partial _{x}\left( B^{3}+\frac{(B^{1})^{2}}{2}-\frac{(B^{0})^{2}B^{1}}{3}%
-\frac{(B^{0})^{4}}{8}+2A^{2}+2A^{0}B^{1}-(B^{0})^{2}A^{0}+(A^{0})^{2}%
\right) .
\end{eqnarray*}

\textbf{Corollary}: A generating function of conservation laws and commuting
flows is given by an auxiliary vertex equation%
\begin{equation}
\partial _{\tau (\zeta )}p(\lambda )=\partial _{x}\left( \frac{s(\zeta
)p(\lambda )}{q(\lambda )-q(\zeta )}\right) .  \label{vert}
\end{equation}%
Indeed, a substitution (\ref{e}) and (\ref{genu}) in the above equation
implies to an identity. Moreover, this generating function can be written
via the sole function $S$%
\begin{equation*}
\partial _{\tau (\zeta )}\frac{\partial _{x}[\ln s(\lambda )]^{\prime }}{%
\partial _{\tau (\lambda )}\ln s(\lambda )}=\partial _{x}\frac{[s(\lambda
)\partial _{x}s^{\prime }(\lambda )-s^{\prime }(\lambda )\partial
_{x}s(\lambda )]s(\zeta )\partial _{\tau (\zeta )}s(\lambda )}{[s(\zeta
)\partial _{x}s(\lambda )-s(\lambda )\partial _{x}s(\zeta )]s(\lambda
)\partial _{\tau (\lambda )}s(\lambda )},
\end{equation*}%
or via infinitely many field variables $q_{k}$%
\begin{equation*}
\partial _{\tau (\zeta )}\frac{\partial _{x}q^{\prime }(\lambda )}{\partial
_{\tau (\lambda )}q(\lambda )}=\partial _{x}\frac{q^{\prime }(\zeta
)\partial _{x}q^{\prime }(\lambda )\partial _{\tau (\zeta )}q(\zeta )}{%
[q(\lambda )-q(\zeta )]\partial _{x}q^{\prime }(\zeta )\partial _{\tau
(\lambda )}q(\lambda )}.
\end{equation*}

\section{Derivation of the Manakov--Santini system}

A substitution of expansions (\ref{apart}) and (\ref{non}) in the \textit{%
second} equations of (\ref{lin}) and (\ref{g}) implies to the first
commuting flow of (\ref{bi})%
\begin{equation*}
\left( 
\begin{array}{c}
A^{k} \\ 
B^{k}%
\end{array}%
\right) _{y}=\left( 
\begin{array}{c}
A^{k+2} \\ 
B^{k+2}%
\end{array}%
\right) _{x}-B^{0}\left( 
\begin{array}{c}
A^{k+1} \\ 
B^{k+1}%
\end{array}%
\right) _{x}-\left( B^{1}-\frac{(B^{0})^{2}}{2}-A^{0}\right) \left( 
\begin{array}{c}
A^{k} \\ 
B^{k}%
\end{array}%
\right) _{x}
\end{equation*}%
\begin{equation}
+(k+1)\left( 
\begin{array}{c}
A^{k} \\ 
B^{k}%
\end{array}%
\right) A_{x}^{0}+k\left( 
\begin{array}{c}
A^{k-1} \\ 
B^{k-1}%
\end{array}%
\right) (A_{x}^{1}-B^{0}A_{x}^{0}),\text{ \ }k=0,1,2,...,  \label{sec}
\end{equation}%
where $c=B^{1}-(B^{0})^{2}/2-A^{0}$.

Let us take first two equations from (\ref{bi}) and the first equation from (%
\ref{sec})%
\begin{equation*}
\left( 
\begin{array}{c}
A^{0} \\ 
B^{0}%
\end{array}%
\right) _{t}=\left( 
\begin{array}{c}
A^{1} \\ 
B^{1}%
\end{array}%
\right) _{x}-B^{0}\left( 
\begin{array}{c}
A^{0} \\ 
B^{0}%
\end{array}%
\right) _{x},
\end{equation*}%
\begin{equation*}
\left( 
\begin{array}{c}
A^{1} \\ 
B^{1}%
\end{array}%
\right) _{t}=\left( 
\begin{array}{c}
A^{2} \\ 
B^{2}%
\end{array}%
\right) _{x}-B^{0}\left( 
\begin{array}{c}
A^{1} \\ 
B^{1}%
\end{array}%
\right) _{x}+\left( 
\begin{array}{c}
A^{0} \\ 
B^{0}%
\end{array}%
\right) A_{x}^{0},
\end{equation*}%
\begin{equation*}
\left( 
\begin{array}{c}
A^{0} \\ 
B^{0}%
\end{array}%
\right) _{y}=\left( 
\begin{array}{c}
A^{2} \\ 
B^{2}%
\end{array}%
\right) _{x}-B^{0}\left( 
\begin{array}{c}
A^{1} \\ 
B^{1}%
\end{array}%
\right) _{x}-\left( B^{1}-\frac{(B^{0})^{2}}{2}-A^{0}\right) \left( 
\begin{array}{c}
A^{0} \\ 
B^{0}%
\end{array}%
\right) _{x}+\left( 
\begin{array}{c}
A^{0} \\ 
B^{0}%
\end{array}%
\right) A_{x}^{0}.
\end{equation*}%
An elimination $A^{2}$ and $B^{2}$ leads to three dimensional four component
quasilinear system of the first order%
\begin{equation*}
\left( 
\begin{array}{c}
A^{0} \\ 
B^{0}%
\end{array}%
\right) _{t}=\left( 
\begin{array}{c}
A^{1} \\ 
B^{1}%
\end{array}%
\right) _{x}-B^{0}\left( 
\begin{array}{c}
A^{0} \\ 
B^{0}%
\end{array}%
\right) _{x},
\end{equation*}%
\begin{equation*}
\left( 
\begin{array}{c}
A^{1} \\ 
B^{1}%
\end{array}%
\right) _{t}=\left( 
\begin{array}{c}
A^{0} \\ 
B^{0}%
\end{array}%
\right) _{y}+\left( B^{1}-\frac{(B^{0})^{2}}{2}-A^{0}\right) \left( 
\begin{array}{c}
A^{0} \\ 
B^{0}%
\end{array}%
\right) _{x}.
\end{equation*}%
A substitution $A^{0}=u,B^{0}=a,A^{1}=b$ and $B^{1}=c+a^{2}/2+u$ implies to (%
\ref{6}).

Thus, we have proved that three dimensional four component quasilinear
system of the first order (\ref{6}) is \textit{equivalent} to commuting
hydrodynamic bi-chains (\ref{bi}) and (\ref{sec}). It means that any
solution (obtained by the method of hydrodynamic reductions, see below) of (%
\ref{6}) can be utilized for a construction of a corresponding solution of
two linear systems (\ref{lin}) and (\ref{osem}); then such a solution can be
used for construction of a solution of commuting hydrodynamic bi-chains (\ref%
{bi}) and (\ref{sec}). And vice versa, any solution of commuting
hydrodynamic bi-chains (\ref{bi}) and (\ref{sec}) implies immediately to a
solution of (\ref{6}).

\section{Semi-Hamiltonian hydrodynamic reductions}

Following approach established in \cite{FK}, all field variables $a,b,c,u$
in (\ref{6}) are considered as functions of $N$ Riemann invariants $\lambda
^{k}(x,t,y)$, where hydrodynamic reductions are commuting semi-Hamiltonian
hydrodynamic type systems%
\begin{equation}
\lambda _{t}^{i}=v^{i}(\mathbf{\lambda })\lambda _{x}^{i},\text{ \ \ \ \ }%
\lambda _{y}^{i}=w^{i}(\mathbf{\lambda })\lambda _{x}^{i},\text{ \ \ }%
i=1,2,...,N  \label{7}
\end{equation}%
for any $N$. Characteristic velocities must satisfy the systems (see \cite%
{Tsar})%
\begin{eqnarray}
\partial _{k}\frac{\partial _{j}v^{i}}{v^{j}-v^{i}} &=&\partial _{j}\frac{%
\partial _{k}v^{i}}{v^{k}-v^{i}},\text{ \ }i\neq j\neq k;  \label{9} \\
&&  \notag \\
\frac{\partial _{k}v^{i}}{v^{k}-v^{i}} &=&\frac{\partial _{k}w^{i}}{%
w^{k}-w^{i}},\text{ \ }i\neq k,  \label{10}
\end{eqnarray}%
where $\partial _{k}\equiv \partial /\partial \lambda ^{k}$. A substitution (%
\ref{7}) in (\ref{6}) implies to the so-called \textquotedblleft dispersion
relation\textquotedblright 
\begin{equation}
w^{i}=v^{i^{2}}+av^{i}-c  \label{11}
\end{equation}%
and two differential relations%
\begin{equation}
\partial _{i}(c+u)=v^{i}\partial _{i}a,\text{ \ \ \ \ }\partial
_{i}b=(v^{i}+a)\partial _{i}u.  \label{12}
\end{equation}%
A substitution (\ref{11}) in (\ref{10}) together with the first equation
from (\ref{12}) implies to%
\begin{equation}
\partial _{i}q^{k}=\frac{\partial _{i}u}{q^{i}-q^{k}},\text{ \ }i\neq k,
\label{13}
\end{equation}%
where we introduce $q^{i}=v^{i}+a$ instead $v^{i}$. Compatibility conditions 
$\partial _{k}(\partial _{i}(c+u))=\partial _{i}(\partial _{k}(c+u))$ and $%
\partial _{k}(\partial _{i}b)=\partial _{i}(\partial _{k}b)$ for each pair
of distinct indices lead to%
\begin{eqnarray}
\partial _{ik}u &=&2\frac{\partial _{i}u\cdot \partial _{k}u}{%
(q^{i}-q^{k})^{2}},  \label{14} \\
&&  \notag \\
\partial _{ik}a &=&\frac{\partial _{i}u\cdot \partial _{k}a+\partial
_{k}u\cdot \partial _{i}a}{(q^{i}-q^{k})^{2}}.  \label{15}
\end{eqnarray}%
System (\ref{13})--(\ref{15}) is an analogue of the Gibbons--Tsarev system
for the dKP equation (see \cite{GT}). Equations (\ref{15}) is a \textit{%
linear} system whose variable coefficients are determined by solutions of 
\textit{sub-system} (\ref{13}), (\ref{14}) which is precisely the
aforementioned Gibbons--Tsarev system. Original Gibbons--Tsarev system (\ref%
{13}), (\ref{14}) has a general solution parameterized by $N$ arbitrary
functions of a single variable. Linear system (\ref{15}) also has a general
solution parameterized by $N$ arbitrary functions of a single variable.
Thus, whole system (\ref{13})--(\ref{15}) has a general solution
parameterized by $2N$ arbitrary functions of a single variable.

Thus, this is a first example (in a literature), where number of arbitrary
functions twice bigger than in an ordinary theory (see \cite{FK}). Let us
give an explanation of this \textit{deviation} here. Three dimensional four
component hydrodynamic type system (\ref{6}) should be written in the form%
\begin{equation}
\left( 
\begin{array}{c}
a \\ 
b \\ 
c \\ 
u%
\end{array}%
\right) _{t}=\left( 
\begin{array}{cccc}
0 & 0 & 1 & 1 \\ 
0 & 0 & 0 & c \\ 
c & -1 & -a & 0 \\ 
0 & 1 & 0 & -a%
\end{array}%
\right) \left( 
\begin{array}{c}
a \\ 
b \\ 
c \\ 
u%
\end{array}%
\right) _{x}+\left( 
\begin{array}{cccc}
0 & 0 & 0 & 0 \\ 
0 & 0 & 0 & 1 \\ 
1 & 0 & 0 & 0 \\ 
0 & 0 & 0 & 0%
\end{array}%
\right) \left( 
\begin{array}{c}
a \\ 
b \\ 
c \\ 
u%
\end{array}%
\right) _{y}.  \label{w}
\end{equation}%
A dispersion relation (see \cite{FK}) for any three dimensional hydrodynamic
type system%
\begin{equation*}
\vec{u}_{t}=\hat{A}\vec{u}_{x}+\hat{B}\vec{u}_{y}
\end{equation*}%
is determined by%
\begin{equation*}
\det [\hat{A}+\xi \hat{B}-\mu \hat{E}]=0,
\end{equation*}%
where $\xi $ and $\mu $ are constants, and $\hat{E}$ is an identity matrix.
In our case,%
\begin{equation}
\det \left( 
\begin{array}{cccc}
-\mu & 0 & 1 & 1 \\ 
0 & -\mu & 0 & c+\xi \\ 
c+\xi & -1 & -\mu -a & 0 \\ 
0 & 1 & 0 & -\mu -a%
\end{array}%
\right) =0  \label{det}
\end{equation}%
implies to the \textit{reducible quartic}%
\begin{equation*}
(\mu ^{2}+a\mu -c-\xi )^{2}=0.
\end{equation*}%
Thus, instead a full algebraic equation of the fourth order (with distinct
roots), hydrodynamic type system (\ref{6}) possesses the degenerated
dispersion relation (\textit{conic}) (cf. (\ref{11}))%
\begin{equation*}
\xi =\mu ^{2}+a\mu -c.
\end{equation*}%
An existence of hydrodynamic reductions (\ref{7}) (integrability conditions (%
\ref{9}) and (\ref{10}) are not necessary!) reduces (\ref{w}) to the form
(cf. (\ref{det}))%
\begin{equation*}
0=\left( 
\begin{array}{cccc}
-v^{i} & 0 & 1 & 1 \\ 
0 & -v^{i} & 0 & c+w^{i} \\ 
c+w^{i} & -1 & -v^{i}-a & 0 \\ 
0 & 1 & 0 & -v^{i}-a%
\end{array}%
\right) \partial _{i}\left( 
\begin{array}{c}
a \\ 
b \\ 
c \\ 
u%
\end{array}%
\right) .
\end{equation*}%
Thus, indeed (see (\ref{11})) $\xi $ has a meaning of the characteristic
velocity $\vec{w}$, and $\mu $ has a meaning of the characteristic velocity $%
\vec{v}$. Such an existence of this \textit{double conic} leads to a \textit{%
double freedom} of solutions for generalized Gibbons--Tsarev system (\ref{13}%
)--(\ref{15}).

An existence of such a double conic leads to the following phenomenon. The
number $N$ of Riemann invariants $\lambda ^{k}$ can be split on two parts.
First $M$ Riemann invariants are branch points of the algebraic curve $%
\lambda =\Lambda (\mathbf{\lambda };q)$ (see below); all other $N-M$ Riemann
invariants are just mark points $r^{n}$ on this algebraic curve $\lambda
=\Lambda (\mathbf{\lambda };q)$. $M$ can run all values from $1$ up to $N$.
Another origin of this phenomenon is connected with an observation that
whole hydrodynamic bi-chain (\ref{bi}) contains two parts, where each of
them is determined by a different linear system (see (\ref{lin}) and (\ref%
{osem})). Moreover, first $M$ Riemann invariants are connected with the $A$%
--part of (\ref{bi}), while other $N-M$ Riemann invariants are connected
with the $B$--part of (\ref{bi}).

Since integrable three dimensional four component hydrodynamic type system (%
\ref{6}) determined by linear system (\ref{lin}) is associated with
commuting hydrodynamic bi-chains (\ref{bi}) and (\ref{sec}), to avoid a lot
of unnecessary computations without loss of generality and for simplicity we
shall consider just first hydrodynamic bi-chain (\ref{bi}) below.

\section{Two species of Riemann invariants}

Following the approach presented in \cite{GT}, let us suppose that all
moments $A^{k}$ are functions of $M$ Riemann invariants $\lambda ^{n}$. In
such a case, the first part of (\ref{bi}) reduces to an infinite series of
recursive relations%
\begin{equation*}
\partial _{i}A^{k+1}=q^{i}\partial _{i}A^{k}-kA^{k-1}\partial _{i}A^{0},%
\text{ \ }k=0,1,2,...,
\end{equation*}%
whose compatibility conditions $\partial _{i}(\partial _{j}A^{k})=\partial
_{j}(\partial _{i}A^{k})$ are nothing else but the original Gibbons--Tsarev
system (\ref{13}), (\ref{14}). Moreover, these Riemann invariants $\lambda
^{i}$ are nothing else but branch points $\lambda ^{i}$ of the Riemann
surface $\lambda =\Lambda (\mathbf{\lambda };q)$, i.e. $\lambda ^{i}=\Lambda
(\mathbf{\lambda };q^{i})$, where $q^{i}(\mathbf{\lambda })$ are solutions
of an algebraic system (see \cite{GT} and \cite{OPS})%
\begin{equation*}
\partial _{q}\Lambda |_{q=q^{i}}=0\text{, \ }i=1,2,...,N.
\end{equation*}%
Indeed, linear system (\ref{lin}) reduces to the so-called L\"{o}wner
equation (see \cite{GT})%
\begin{equation}
\partial _{i}\Lambda =\frac{\partial _{i}u}{q-q^{i}}\partial _{q}\Lambda ,
\label{loew}
\end{equation}%
whose compatibility conditions $\partial _{i}(\partial _{k}\Lambda
)=\partial _{k}(\partial _{i}\Lambda )$ imply to original Gibbons--Tsarev
system (\ref{13}), (\ref{14}). Thus, hydrodynamic bi-chain (\ref{bi})
reduces to a \textit{composition} of the integrable hydrodynamic chain%
\begin{equation}
B_{t}^{k}=B_{x}^{k+1}-B^{0}B_{x}^{k}+kB^{k-1}u_{x},\text{ \ }k=0,1,2,...
\label{chain}
\end{equation}%
and the semi-Hamiltonian hydrodynamic type system (see the first equation in
(\ref{7}))%
\begin{equation}
\lambda _{t}^{i}=(q^{i}(\mathbf{\lambda })-B^{0})\lambda _{x}^{i},\text{ \ \ 
}i=1,2,...,M,  \label{add}
\end{equation}%
where functions $u$ and $q^{i}(\mathbf{\lambda })$ satisfy original
Gibbons--Tsarev system (\ref{13}), (\ref{14}). Nevertheless, linear system (%
\ref{15}) cannot be derived from (\ref{lin}), because $B^{0}$ is not a
function of Riemann invariants $\lambda ^{i}$. However, if we consider
hydrodynamic reductions of hydrodynamic chain (\ref{chain}), i.e. if we
suppose that $B^{0}$ is a function of Riemann invariants $\lambda ^{i}$,
then all higher moments $B^{k}$ must be functions of Riemann invariants too.
Then compatibility conditions $\partial _{i}(\partial _{j}B^{k})=\partial
_{j}(\partial _{i}B^{k})$ lead to (\ref{15}), where%
\begin{equation*}
\partial _{i}B^{k+1}=q^{i}\partial _{i}B^{k}-kB^{k-1}\partial _{i}u.
\end{equation*}

A most interesting consequence of aforementioned degeneracy described in the
previous Section is an existence of \textit{extra} Riemann invariants which
are just \textit{mark points} on already determined Riemann surface $\lambda
=\Lambda (\mathbf{\lambda };q)$. Indeed, let us introduce $N-M$ mark points $%
r^{k}=\Lambda (\mathbf{\lambda };\tilde{q}^{k})$ such that each inverse
function $\tilde{q}^{j}(\mathbf{\lambda },r^{j})$ is a solution of the L\"{o}%
wner equation (see (\ref{loew}), where we consider an inverse function $q(%
\mathbf{\lambda };\lambda )$)%
\begin{equation}
\partial _{i}q=\frac{\partial _{i}u}{q^{i}-q},\text{ \ }i=1,2,...,M.
\label{new}
\end{equation}%
A general solution of this overdetermined system (i.e. the compatibility
conditions $\partial _{i}(\partial _{k}\tilde{q}^{j})=\partial _{k}(\partial
_{i}\tilde{q}^{j})$ must be fulfilled due to original Gibbons--Tsarev system
(\ref{13}), (\ref{14}))%
\begin{equation}
\partial _{i}\tilde{q}^{j}=\frac{\partial _{i}u}{q^{i}-\tilde{q}^{j}},\text{
\ }i=1,2,...,M,\text{ \ }j=1,2,...,N-M  \label{qwa}
\end{equation}%
depends on $M$ arbitrary functions of a single variable (because this is a
system of the first order in partial derivatives with respect to $M$
independent variables $\lambda ^{i}$) and other $N-M$ arbitrary functions of
a single variable (because each Riemann invariant $r^{k}$ as a mark point on
the Riemann surface $\lambda =\Lambda (\mathbf{\lambda };q)$ is determined
up to an arbitrary transformation $r^{k}\rightarrow R_{k}(r^{k})$. Let us
remind that a solution $\lambda $ of linear system (\ref{lin}) is determined
up to an arbitrary transformation $\lambda \rightarrow \tilde{\lambda}%
(\lambda )$). Then let us introduce the functions $b_{j}(\mathbf{\lambda }%
,r^{j})$ determined by their first derivatives (the compatibility conditions 
$\partial _{i}(\partial _{k}b_{j})=\partial _{k}(\partial _{i}b_{j})$ are
fulfilled due to original Gibbons--Tsarev system (\ref{13}), (\ref{14}))%
\begin{equation}
\partial _{i}c_{j}(\mathbf{\lambda },r^{j})=\frac{\partial _{i}u}{(q^{i}-%
\tilde{q}^{j})^{2}}.  \label{comp}
\end{equation}%
such that%
\begin{equation}
B^{0}(\mathbf{\lambda },\mathbf{r})=\sum_{j=1}^{N-M}\int \exp c_{j}(\mathbf{%
\lambda },r^{j})dr^{j}  \label{bo}
\end{equation}%
satisfies overdetermined system (\ref{15}). Whole overdetermined system (\ref%
{15}), (\ref{qwa}), (\ref{comp}) can be rewritten as a quasilinear system of
the first order on unknown functions $\tilde{q}^{j},\partial _{i}B^{0}$ and $%
\tilde{\partial}_{j}B^{0}=\exp c_{j}$ (here and below in this Section $%
\tilde{\partial}_{j}=\partial /\partial r^{j}$). Thus, its general solution
is parameterized by $N$ arbitrary functions of a single variable. It means,
a general solution of complete overdetermined system (\ref{13})--(\ref{15}),
(\ref{qwa}), (\ref{comp}) is parameterized by $N+M$ arbitrary functions of a
single variable.

Let us introduce $N$ component hydrodynamic type system (see (\ref{add}))%
\begin{equation}
\lambda _{t}^{i}=(q^{i}(\mathbf{\lambda })-B^{0})\lambda _{x}^{i},\text{ \ \ 
}r_{t}^{j}=(\tilde{q}^{j}(\mathbf{\lambda },r^{j})-B^{0})r_{x}^{j}.
\label{extra}
\end{equation}%
An integrability condition (i.e. semi-Hamiltonian property (\ref{9})) is
fulfilled automatically due to (\ref{13})--(\ref{15}), (\ref{qwa}), (\ref%
{comp}).

\textbf{Theorem}: \textit{Hydrodynamic bi-chain} (\ref{bi}) \textit{%
possesses infinitely many }$N$\textit{\ component hydrodynamic reductions }(%
\ref{extra}) \textit{determined by the \textbf{extended} Gibbons--Tsarev
system} (\ref{13})--(\ref{15}), (\ref{qwa}), (\ref{comp})\textit{. Then all
higher moments }$B^{k}$\textit{\ can be reconstructed iteratively in
quadratures}%
\begin{equation}
dB^{k+1}=\sum_{i=1}^{M}[q^{i}\partial _{i}B^{k}-kB^{k-1}\partial
_{i}u]d\lambda ^{i}+\sum_{j=1}^{N-M}\tilde{q}^{j}\tilde{\partial}%
_{j}B^{k}dr^{j},\text{ \ }k=0,1,2,...  \label{ful}
\end{equation}

\textbf{Proof}: Since all moments $B^{k}$ depend simultaneously on both
species of Riemann invariants $\lambda ^{i}$ and $r^{j}$, then the $B$-part
of (\ref{bi}) reduces to the second equation in (\ref{extra}), where (see (%
\ref{ful}))%
\begin{equation}
\partial _{i}B^{k+1}=q^{i}\partial _{i}B^{k}-kB^{k-1}\partial _{i}u,\text{ \
\ }\tilde{\partial}_{j}B^{k+1}=\tilde{q}^{j}\tilde{\partial}_{j}B^{k}.
\label{aha}
\end{equation}%
The compatibility conditions $\tilde{\partial}_{j}(\tilde{q}^{i}\tilde{%
\partial}_{i}B^{k})=\tilde{\partial}_{i}(\tilde{q}^{j}\tilde{\partial}%
_{j}B^{k}),\partial _{i}(\tilde{q}^{j}\tilde{\partial}_{j}B^{k})=\tilde{%
\partial}_{j}(q^{i}\partial _{i}B^{k}-kB^{k-1}\partial _{i}u)$, $\partial
_{j}(q^{i}\partial _{i}B^{k}-kB^{k-1}\partial _{i}u)=\partial
_{i}(q^{j}\partial _{j}B^{k}-kB^{k-1}\partial _{j}u)$ lead to the system ($%
i\neq j$ in the first and third equations below)%
\begin{eqnarray}
\tilde{\partial}_{i}\tilde{\partial}_{j}B^{k} &=&0,\text{ \ \ \ \ \ }(q^{i}-%
\tilde{q}^{j})\partial _{i}\tilde{\partial}_{j}B^{k}=\partial _{i}\tilde{q}%
^{j}\cdot \tilde{\partial}_{j}B^{k}+k\partial _{i}u\cdot \tilde{\partial}%
_{j}B^{k-1},  \notag \\
&&  \label{l} \\
(q^{i}-q^{j})^{2}\partial _{i}\partial _{j}B^{k} &=&\partial _{i}u\cdot
\partial _{j}B^{k}+\partial _{j}u\cdot \partial
_{i}B^{k}+k(q^{i}-q^{j})(\partial _{i}u\cdot \partial _{j}B^{k-1}-\partial
_{j}u\cdot \partial _{i}B^{k-1}).  \notag
\end{eqnarray}%
The second equation in (\ref{aha}) can be written in the common form%
\begin{equation}
\tilde{\partial}_{j}B^{k}=(\tilde{q}^{j})^{k}\tilde{\partial}_{j}B^{0},\text{
\ }k=0,1,2,...  \label{mak}
\end{equation}%
A substitution (\ref{mak}) in the first two equations of (\ref{l}) leads to (%
\ref{qwa}), (\ref{comp}) and (\ref{bo}), where each function $\tilde{q}^{j}$
depends on all Riemann invariants $\lambda ^{i}$ and just one Riemann
invariant $r^{j}$. The third equation in (\ref{l}) coincides with (\ref{15})
for $k=0$. The compatibility conditions $\partial _{i}(\partial
_{jl}B^{k})=\partial _{j}(\partial _{il}B^{k})$ can be verified by the
induction principle (taking into account the first equation in (\ref{aha}).
Precisely, such a computation was made for $A$-part of hydrodynamic bi-chain
(\ref{bi}), i.e. for Benney hydrodynamic chain (\ref{raz}) in \cite{GT}.

Correspondingly, linear problem (\ref{osem}) determining the $B$-parts of
hydrodynamic bi-chains (\ref{bi}) and (\ref{sec}) reduces to (cf. (\ref{loew}%
))%
\begin{equation}
\partial _{i}F=\frac{\partial _{i}a}{q^{i}-q},\text{ \ \ \ }\tilde{\partial}%
_{j}F=\frac{\tilde{\partial}_{j}a}{\tilde{q}^{j}-q},  \label{red}
\end{equation}%
where $F(\mathbf{\lambda },\mathbf{r})$ is used instead of $f(x,t,y,q)$ for
hydrodynamic reductions (\ref{7}). The dependence $\tilde{q}^{j}(\mathbf{%
\lambda },r^{j})$ and (\ref{bo}) follow from the compatibility condition $%
\tilde{\partial}_{i}(\tilde{\partial}_{j}F)=\tilde{\partial}_{j}(\tilde{%
\partial}_{i}F)$. The compatibility condition $\partial _{i}(\tilde{\partial}%
_{j}F)=\tilde{\partial}_{j}(\partial _{i}F)$ leads to (\ref{qwa}) and (\ref%
{comp}). The compatibility condition $\partial _{i}(\partial _{j}F)=\partial
_{j}(\partial _{i}F)$ satisfies automatically due to original
Gibbons--Tsarev system (\ref{13}), (\ref{14}). Similar computations can be
repeated for generation function of conservation laws (\ref{geni}). Then the
generating function of conservation law densities $p$ can be found in
quadratures%
\begin{equation}
d\ln p=\sum_{i=1}^{M}\left( \frac{\partial _{i}u}{(q-q^{i})^{2}}+\frac{%
\partial _{i}a}{q-q^{i}}\right) d\lambda ^{i}+\sum_{j=1}^{N-M}\frac{\tilde{%
\partial}_{j}a}{q-\tilde{q}^{j}}dr^{j}.  \label{de}
\end{equation}

It is well known that generating function of conservation laws (\ref{geni})
is associated with the so-called \textquotedblleft linearly
degenerate\textquotedblright\ hydrodynamic chains and their linearly
degenerate hydrodynamic reductions and (see \cite{MaksEps} and Section 9 in 
\cite{algebra}). First such an example was the Whitham equations of the
Korteweg de Vries equation (see \cite{FFM}). More general theory was
suggested in \cite{Krich}. Theory of linearly degenerate hydrodynamic type
systems is presented in \cite{EKPZ}, \cite{Ferlin}, \cite{Makslin}.
Following \cite{FFM} and \cite{Krich}, let us introduce the so-called
\textquotedblleft quasi-momentum\textquotedblright\ differential\ $%
dP=pd\lambda $ and the so-called \textquotedblleft
quasi-energy\textquotedblright\ differential\ $dQ=(q-a)pd\lambda $. Then
characteristic velocities are given by%
\begin{equation*}
\tilde{q}^{j}(\mathbf{\lambda },r^{j})-B^{0}(\mathbf{\lambda },\mathbf{r})=%
\frac{dQ}{dP}|_{\lambda =r^{j}},
\end{equation*}%
where the differentials of \textquotedblleft quasi-energy\textquotedblright\
and \textquotedblleft quasi-momentum\textquotedblright\ possess similar
singularities on the Riemann surface $\lambda =\Lambda (\mathbf{\lambda };q)$
at vicinities of mark points $\lambda =r^{i}$.

Generating function of conservation laws and commuting flows (\ref{vert})
can be written in similar form. In such a case, the generating function of
\textquotedblleft quasi-energy\textquotedblright\ differentials is given by%
\begin{equation*}
d\tilde{Q}(\lambda ,\zeta )=W(\lambda ,\zeta )p(\lambda )d\lambda \equiv 
\frac{\exp F(\zeta )}{q(\lambda )-q(\zeta )}p(\lambda )d\lambda .
\end{equation*}%
Corresponding generating function of commuting hydrodynamic type systems
(see (\ref{extra}))%
\begin{equation*}
\lambda _{\tau (\zeta )}^{i}=W^{i}(\mathbf{\lambda };\zeta )\lambda _{x}^{i},%
\text{ \ \ }r_{\tau (\zeta )}^{j}=\tilde{W}^{j}(\mathbf{\lambda ,r};\zeta
)r_{x}^{j}
\end{equation*}%
is given by (replacing $q(\lambda )$ in the expression for $W(\lambda ,\zeta
)$ by $q^{i}$ and $\tilde{q}^{j}$, respectively)%
\begin{equation}
\lambda _{\tau (\zeta )}^{i}=\frac{\exp F(\zeta )}{q^{i}-q(\zeta )}\lambda
_{x}^{i},\text{ \ \ }r_{\tau (\zeta )}^{j}=\frac{\exp F(\zeta )}{\tilde{q}%
^{j}-q(\zeta )}r_{x}^{j}.  \label{genn}
\end{equation}%
This hydrodynamic type system possesses generating function (\ref{vert}). In
such a case (cf. Section 9 in \cite{algebra}),%
\begin{equation*}
d\ln p=\sum_{i=1}^{M}\frac{\partial _{i}W}{W^{i}-W}d\lambda
^{i}+\sum_{j=1}^{N-M}\frac{\tilde{\partial}_{j}W}{\tilde{W}^{j}-W}dr^{j}
\end{equation*}%
must coincide with (\ref{de}). Indeed, a direct substitution of expressions $%
W(\lambda ,\zeta ),W^{i}(\mathbf{\lambda };\zeta )$ and $\tilde{W}^{j}(%
\mathbf{\lambda ,r};\zeta )$ implies to (\ref{de}). Moreover, hydrodynamic
type systems (\ref{extra}) and (\ref{genn}) commute to each other (see
Tsarev's condition (\ref{10})). It means, that any hydrodynamic reduction (%
\ref{extra}) satisfying extended Gibbons--Tsarev system (\ref{13})--(\ref{15}%
), (\ref{qwa}), (\ref{comp}) is compatible with the whole Manakov--Santini
hierarchy described in Section 3.

\textbf{Remark}: Semi-Hamiltonian property (\ref{9}) and commutativity
condition (\ref{10}) for hydrodynamic type systems which are hydrodynamic
reductions of linearly degenerate hydrodynamic chains can be simplified (see
Section 9 in \cite{algebra}) to the form, respectively%
\begin{equation*}
\partial _{i}\frac{\partial _{k}V}{V^{k}-V}=\partial _{k}\frac{\partial _{i}V%
}{V^{i}-V},\text{\ }i\neq k;\text{ \ \ }\frac{\partial _{k}V}{V^{k}-V}=\frac{%
\partial _{k}W}{W^{k}-W},
\end{equation*}%
where $dP=p(\lambda )d\lambda $ is a differential of \textquotedblleft
quasi-momentum\textquotedblright , $dQ=V(\lambda )p(\lambda )d\lambda $ is a
differential of \textquotedblleft quasi-energy\textquotedblright\ and $d%
\tilde{Q}=W(\lambda ,\zeta )p(\lambda )d\lambda $ is a generating function
of \textquotedblleft quasi-energy\textquotedblright\ differentials. In the
case of two species of Riemann invariants (see, for instance, (\ref{extra})
and (\ref{genn})), the above formulas must be written in the form%
\begin{equation*}
\partial _{i}\frac{\partial _{k}V}{V^{k}-V}=\partial _{k}\frac{\partial _{i}V%
}{V^{i}-V},\text{ \ }\tilde{\partial}_{i}\frac{\tilde{\partial}_{k}V}{\tilde{%
V}^{k}-V}=\tilde{\partial}_{k}\frac{\tilde{\partial}_{i}V}{\tilde{V}^{i}-V},%
\text{ \ }i\neq k;\text{ \ \ }\tilde{\partial}_{i}\frac{\partial _{k}V}{%
V^{k}-V}=\partial _{k}\frac{\tilde{\partial}_{i}V}{\tilde{V}^{i}-V},
\end{equation*}%
\begin{equation*}
\frac{\partial _{k}V}{V^{k}-V}=\frac{\partial _{k}W}{W^{k}-W},\text{ \ \ \ }%
\frac{\tilde{\partial}_{k}V}{\tilde{V}^{k}-V}=\frac{\tilde{\partial}_{k}W}{%
\tilde{W}^{k}-W}.
\end{equation*}

\section{Symmetric hydrodynamic reductions}

Benney hydrodynamic chain (\ref{raz}) possesses the special hydrodynamic
reduction (see \cite{MaksHam})%
\begin{equation*}
a_{t}^{i}=\left( \frac{(a^{i})^{2}}{2}+A^{0}\right) _{x},
\end{equation*}%
where all moments are determined by the choice ($\epsilon _{i}$ are
constants)%
\begin{equation*}
A^{k}=\frac{1}{k+1}\sum_{i=1}^{M}\epsilon _{i}(a^{i})^{k+1},\text{ \ }%
k=0,1,2,...;\text{ \ \ \ }\sum_{i=1}^{M}\epsilon _{i}=0.
\end{equation*}%
Let us extend this decomposition on both species of moments ($\epsilon _{i}$
and $\gamma _{j}$ are constants)%
\begin{equation}
A^{k}=\frac{1}{k+1}\sum_{i=1}^{M}\epsilon _{i}(a^{i})^{k+1},\text{ \ \ }%
B^{k}=\frac{1}{k+1}\sum_{j=1}^{N-M}\gamma _{j}(b^{j})^{k+1},\text{ \ }%
k=0,1,2,...  \label{k}
\end{equation}%
Then bi-chain (\ref{bi}) reduces to the hydrodynamic type system%
\begin{equation}
a_{t}^{i}=(a^{i}-B^{0})a_{x}^{i}+A_{x}^{0},\text{ \ \ \ }%
b_{t}^{j}=(b^{j}-B^{0})b_{x}^{j}+A_{x}^{0},  \label{sys}
\end{equation}%
where%
\begin{equation*}
\sum_{i=1}^{M}\epsilon _{i}=\sum_{j=1}^{N-M}\gamma _{j}=0.
\end{equation*}

In a contrary with semi-Hamiltonian hydrodynamic reductions (see (\ref{chain}%
) and (\ref{add})) of the $A$-part of bi-chain (\ref{bi}), the moment
decomposition (see (\ref{k}))%
\begin{equation*}
B^{k}=\frac{1}{k+1}\sum_{j=1}^{N-M}\gamma _{j}(b^{j})^{k+1},\text{ \ }%
k=0,1,2,...
\end{equation*}%
for the $B$-part of (\ref{bi}) leads to the hydrodynamic chain%
\begin{equation*}
A_{t}^{k}=A_{x}^{k+1}-B^{0}A_{x}^{k}+kA^{k-1}A_{x}^{0},\text{ \ }k=0,1,2,...
\end{equation*}%
equipped by the \textit{non-diagonalizable} (i.e. non semi-Hamiltonian!)
hydrodynamic type system (cf. (\ref{sys}))%
\begin{equation}
b_{t}^{j}=(b^{j}-B^{0})b_{x}^{j}+A_{x}^{0},\text{ \ \ }j=1,2,...,N-M.
\label{c}
\end{equation}%
Indeed, Riemann invariants $r^{k}(\mathbf{b})$ cannot exist, because the
last term $A_{x}^{0}$ cannot be eliminated by \textit{any} point
transformations. Moreover, in a general case, the $A$-part of (\ref{bi})
reduces to the $M$ component hydrodynamic type system%
\begin{equation}
a_{t}^{i}=(a^{i}-B^{0})a_{x}^{i}+A_{x}^{0},\text{ \ \ }i=1,2,...,M,
\label{aj}
\end{equation}%
where all moments $A^{k}$ depend on field variables $a^{n}$ only, and the
function $u=A^{0}$ satisfies the original Gibbons--Tsarev system (see \cite%
{algebra})%
\begin{equation}
(a^{i}-a^{k})u_{ik}+u_{i}\delta u_{k}-u_{k}\delta u_{i}=0,\text{ \ }i\neq k,
\label{unik}
\end{equation}%
where $\delta =\Sigma \partial /\partial a^{m}$ is a shift operator, $%
u_{i}=\partial u/\partial a^{i},u_{ik}=\partial ^{2}u/\partial a^{i}\partial
a^{k}$, all higher moments $A^{k}$ can be reconstructed iteratively (here
and below $\partial _{i}A^{k}=\partial A^{k}/\partial a^{i}$) in quadratures%
\begin{equation*}
dA^{k+1}=\sum_{i=1}^{M}[a^{i}\partial _{i}A^{k}+(\delta
A^{k}-kA^{k-1})\partial _{i}A^{0}]da^{i},\text{ \ }k=0,1,2,...
\end{equation*}%
Linear system (\ref{lin}) reduces to the L\"{o}wner equation (cf. (\ref{loew}%
); more detail in \cite{algebra})%
\begin{equation}
\partial _{i}\Lambda +\frac{\partial _{i}u}{a^{i}-q}\left( 1+\sum_{m=1}^{M}%
\frac{\partial _{m}u}{a^{m}-q}\right) ^{-1}\partial _{q}\Lambda =0,
\label{lambda}
\end{equation}%
written via field variables $a^{k}$, where the equation of Riemann surface $%
\lambda =\Lambda (\mathbf{a};q)$. The compatibility conditions $\partial
_{i}(\partial _{j}\Lambda )=\partial _{j}(\partial _{i}\Lambda )$ as well as
the compatibility conditions $\partial _{i}[a^{j}\partial _{j}A^{k}+(\delta
A^{k}-kA^{k-1})\partial _{j}A^{0}]=\partial _{j}[a^{i}\partial
_{i}A^{k}+(\delta A^{k}-kA^{k-1})\partial _{i}A^{0}]$ lead to (\ref{unik}).

In a contrary, the $B$-part of (\ref{bi}) reduces to (\ref{c}) if all
moments $B^{k}$ depend on field variables $b^{n}$ only. It means, that in a
general case, moments $B^{k}$ must depend on two species of field variables $%
a^{n}$ and $b^{m}$ simultaneously (as well as on two species of Riemann
invariants $\lambda ^{i}$ and $r^{j}$). Just in such a case, hydrodynamic
type system (\ref{c}), (\ref{aj}) can be semi-Hamiltonian. Indeed, the $B$%
-part of (\ref{bi})%
\begin{equation*}
B_{t}^{k}=B_{x}^{k+1}-B^{0}B_{x}^{k}+kB^{k-1}A_{x}^{0},\text{ \ }k=0,1,2,...
\end{equation*}%
reduces to (\ref{c}) and (\ref{aj}), if (here and below $\tilde{\partial}%
_{i}=\partial /\partial b^{i},\tilde{\delta}=\Sigma \partial /\partial b^{m}$%
)%
\begin{equation}
\partial _{i}B^{k+1}=a^{i}\partial _{i}B^{k}+(\delta B^{k}+\tilde{\delta}%
B^{k}-kB^{k-1})\partial _{i}A^{0},\text{ \ \ }\tilde{\partial}%
_{j}B^{k+1}=b^{j}\tilde{\partial}_{j}B^{k}.  \label{ch}
\end{equation}%
The last equation can be written in the common form (cf. (\ref{mak}))%
\begin{equation*}
\tilde{\partial}_{j}B^{k}=(b^{j})^{k}\tilde{\partial}_{j}B^{0},\text{ \ }%
k=0,1,2,...
\end{equation*}

The compatibility conditions $\tilde{\partial}_{j}[a^{i}\partial
_{i}B^{k}+(\delta B^{k}+\tilde{\delta}B^{k}-kB^{k-1})\partial
_{i}A^{0}]=\partial _{i}[b^{j}\tilde{\partial}_{j}B^{k}]$, $\partial
_{j}[a^{i}\partial _{i}B^{k}+(\delta B^{k}+\tilde{\delta}B^{k}-kB^{k-1})%
\partial _{i}A^{0}]=\partial _{i}[a^{j}\partial _{j}B^{k}+(\delta B^{k}+%
\tilde{\delta}B^{k}-kB^{k-1})\partial _{j}A^{0}]$ and $\tilde{\partial}%
_{i}(b^{j}\tilde{\partial}_{j}B^{k})=\tilde{\partial}_{j}(b^{i}\tilde{%
\partial}_{i}B^{k})$ lead to the system (cf. (\ref{l}); $i\neq j$ in the
first and third equations below)%
\begin{equation}
\tilde{\partial}_{i}\tilde{\partial}_{j}B^{0}=0,\text{ \ \ }%
(a^{i}-b^{j})\partial _{i}\tilde{\partial}_{j}B^{0}+\partial _{i}A^{0}\cdot 
\tilde{\partial}_{j}(\delta B^{0}+\tilde{\delta}B^{0})=0,  \label{dom}
\end{equation}%
\begin{equation*}
(a^{i}-a^{j})\partial _{i}\partial _{j}B^{k}+\partial _{i}A^{0}\cdot
\partial _{j}(\delta B^{k}+\tilde{\delta}B^{k}-kB^{k-1})-\partial
_{j}A^{0}\cdot \partial _{i}(\delta B^{k}+\tilde{\delta}B^{k}-kB^{k-1})=0.
\end{equation*}%
The second equation in (\ref{dom}) reduces to%
\begin{equation}
\partial _{i}\tilde{\lambda}_{j}+\frac{\partial _{i}u}{a^{i}-b^{j}}\left(
1+\sum_{m=1}^{M}\frac{\partial _{m}u}{a^{m}-b^{j}}\right) ^{-1}\tilde{%
\partial}_{j}\tilde{\lambda}_{j}=0,  \label{reda}
\end{equation}%
which can be also obtained from (\ref{lambda}) formally replacing $\Lambda $
on $N-M$ functions $\tilde{\lambda}_{j}$ and correspondingly $q$ on $N-M$
field variables $b^{j}$. In such a case, a solution of the first equation in
(\ref{dom}) is given by (cf. (\ref{bo}))%
\begin{equation}
B^{0}=\sum_{j=1}^{N-M}\int \tilde{\lambda}_{j}(\mathbf{a};b^{j})db^{j}
\label{bio}
\end{equation}%
A dependence of $B^{0}$ with respect to field variables $a^{k}$ is given by
the last equation in (\ref{dom})%
\begin{equation}
(a^{i}-a^{j})\partial _{i}\partial _{j}B^{0}+\partial _{i}A^{0}\cdot
\partial _{j}(\delta B^{0}+\tilde{\delta}B^{0})-\partial _{j}A^{0}\cdot
\partial _{i}(\delta B^{0}+\tilde{\delta}B^{0})=0,\text{ \ }i\neq j,
\label{y}
\end{equation}%
while all other higher expressions $B^{k}(\mathbf{a},\mathbf{b})$ can be
found iteratively (see (\ref{ch}))%
\begin{equation*}
dB^{k+1}=\sum_{i=1}^{M}[a^{i}\partial _{i}B^{k}+(\delta B^{k}+\tilde{\delta}%
B^{k}-kB^{k-1})\partial _{i}A^{0}]da^{i}+\sum_{j=1}^{N-M}b^{j}\tilde{\partial%
}_{j}B^{k}db^{j},\text{ \ }k=0,1,2,...
\end{equation*}

The L\"{o}wner equation (\ref{lambda}) under the inverse transformation $%
\Lambda (\mathbf{a};q)\rightarrow q(\mathbf{a};\lambda )$ (cf. (\ref{loew})
and (\ref{new})) reduces to the form%
\begin{equation}
\partial _{i}q=\frac{\partial _{i}u}{a^{i}-q}\left( 1+\sum_{m=1}^{M}\frac{%
\partial _{m}u}{a^{m}-q}\right) ^{-1}  \label{dec}
\end{equation}%
and the generating function of conservation law densities $p$ (see (\ref%
{geni})) can be found in quadratures (cf. (\ref{de}))%
\begin{equation*}
d\ln p=\sum_{m=1}^{M}\partial _{m}(q-a)\cdot d\ln (q-a^{m})-\sum_{n=1}^{N-M}%
\tilde{\partial}_{n}a\cdot d\ln (q-b^{n}),
\end{equation*}%
while (cf. (\ref{red})) the function $F(\mathbf{a},\mathbf{b})$ is given by
another quadrature (see (\ref{exp}) and (\ref{genu}))%
\begin{equation*}
dF=\sum_{m=1}^{M}\partial _{m}a\cdot d\ln (q-a^{m})+\sum_{n=1}^{N-M}\tilde{%
\partial}_{n}a\cdot d\ln (q-b^{n}).
\end{equation*}

\textbf{Remark}: System (\ref{y}) automatically satisfies (for any solution $%
A^{0}(\mathbf{a})$ of system (\ref{unik})) if%
\begin{equation}
B^{0}=\sum_{m=1}^{M}\gamma _{m}a^{m}+\sum_{n=1}^{N-M}\beta _{n}b^{n},
\label{simple}
\end{equation}%
where $\gamma _{m}$ and $\beta _{n}$ are arbitrary constants. In such a case%
\begin{equation*}
F=\sum_{m=1}^{M}\gamma _{m}\ln (q-a^{m})+\sum_{n=1}^{N-M}\beta _{n}\ln
(q-b^{n}),\text{ \ \ \ }p=\frac{1}{\lambda _{q}}\overset{M}{\underset{m=1}{%
\dprod }}(q-a^{m})^{-\gamma _{m}}\overset{N-M}{\underset{n=1}{\dprod }}%
(q-b^{n})^{-\beta _{n}}.
\end{equation*}

All moments $A^{k}$ can be expressed via field variables $a^{i}$ as well as
Riemann invariants $\lambda ^{j}$. Thus, each field variable $a^{i}$ is a
function of all Riemann invariants $\lambda ^{n}$, and vice versa each
Riemann invariant $\lambda ^{i}$ is a function of all Riemann invariants $%
a^{n}$. Since all moments $B^{k}$ are functions of both species of field
variables $a^{i}$ and $b^{j}$ as well as both species of Riemann invariants $%
\lambda ^{i}$ and $r^{j}$, finally, we need to find a dependence of field
variables $b^{k}$ via Riemann invariants $\lambda ^{i}$ and $r^{j}$. Taking
into account (\ref{extra}), (\ref{c}) and (\ref{aj}) imply to%
\begin{equation}
(\tilde{q}^{j}-b^{n})\cdot \tilde{\partial}_{j}b^{n}=0,\text{ \ }%
j,n=1,2,...,N-M  \label{relat}
\end{equation}%
and ($\partial _{i}=\partial /\partial \lambda ^{i},\tilde{\partial}%
_{j}=\partial /\partial r^{j}$ in (\ref{relat}) and below in this Section)%
\begin{equation*}
\partial _{i}a^{k}=\frac{\partial _{i}u}{q^{i}-a^{k}},\text{ \ \ \ }\partial
_{i}b^{n}=\frac{\partial _{i}u}{q^{i}-b^{n}},\text{ \ }i,k=1,2,...,M,
\end{equation*}%
which can be obtained directly from (\ref{new}) replacing $q$ by $a^{k}$ and 
$b^{n}$, respectively. Relations (\ref{relat}) imply to the simple
dependence $b^{j}=\tilde{q}^{j}(\mathbf{\lambda },r^{j})$. Since each
Riemann invariant $r^{i}$ is a mark point on the Riemann surface $\lambda
=\Lambda (\mathbf{\lambda };q)$, i.e. $r^{j}=\Lambda (\mathbf{\lambda };%
\tilde{q}^{j})$, we conclude that equations (\ref{aj}) are connected with (%
\ref{add}), while equations (\ref{c}) reduce to the diagonal form%
\begin{equation*}
r_{t}^{j}=(b^{j}-a)r_{x}^{j},
\end{equation*}%
where Riemann invariants $r^{j}=\Lambda (\mathbf{\lambda }%
;b^{j}),j=1,2,...,N-M$. Thus (see (\ref{reda})), $\tilde{\lambda}_{j}(%
\mathbf{a};b^{j})=R_{j}(r^{j})$, where $R_{j}(r^{j})$ are arbitrary
functions, because Riemann invariants are determined up to an arbitrary
transformation $r^{j}\rightarrow R_{j}(r^{j})$.

In the next Section, explicit hydrodynamic reductions are considered. A
relationship between field variables $a^{k},b^{n}$ and Riemann invariants $%
\lambda ^{i},r^{j}$ is investigated in detail.

\section{Explicit hydrodynamic reductions}

A general case contains an arbitrary number of field variables of two
species $a^{k}$ and $b^{n}$ in $N$ component hydrodynamic type system (\ref%
{sys}). The most general reduction of $\lambda (p)$ known recently (see \cite%
{bk}, \cite{Krich}, \cite{strach}) is a combination of rational and
logarithmic functions with respect to $p$, i.e.%
\begin{eqnarray*}
\lambda  &=&\frac{p^{N+1}}{N+1}%
+a_{0}^{(0)}p^{N-1}+a_{1}^{(0)}p^{N-2}+...+a_{N-1}^{(0)} \\
&& \\
&&+\sum_{k=1}^{K}\left[ \sum_{n=1}^{N_{k}}\frac{a_{k}^{(1)}}{\left(
p-a_{k}^{(2)}\right) ^{n}}+\epsilon _{k}\ln \left( p-a_{k}^{(2)}\right) %
\right] +\sum_{m=1}^{M}\delta _{k}\ln \left( p-a_{k}^{(3)}\right) ,
\end{eqnarray*}%
where $\epsilon _{k}$ and $\delta _{n}$ are constants, $a_{k}^{(n)}$ are
functions. The approach allowing to extract more complicated reductions is
presented in \cite{maksbenney}. In this Section a simplest case is
considered.

$N$ parametric solution of original Gibbons--Tsarev system (\ref{unik})
given by (see \cite{MaksHam}; here all constants $\epsilon _{i}$ are
independent, no such a constraint $\Sigma \epsilon _{m}=0$ in general)%
\begin{equation}
u=\sum_{m=1}^{M}\epsilon _{m}a^{m}  \label{anc}
\end{equation}%
possesses to reconstruct an equation of the Riemann surface $\lambda
=\Lambda (\mathbf{a};q)$. A substitution (\ref{anc}) in (\ref{lambda}) leads
to the so-called waterbag reduction (see \cite{Kodama}, \cite{GT} and \cite%
{MaksHam})%
\begin{equation}
\lambda =q-\sum_{m=1}^{M}\epsilon _{m}\ln (q-a^{m}).  \label{water}
\end{equation}%
The Riemann invariants $\lambda ^{i}(\mathbf{a})$ are branch points of the
Riemann surface determined by (\ref{water}), i.e. the condition ($\lambda
_{q}=0$)%
\begin{equation*}
1=\sum_{m=1}^{M}\frac{\epsilon _{m}}{q^{i}-a^{m}}
\end{equation*}%
leads to $M$ expressions $q^{i}(\mathbf{a})$ as well as to $M$ inverse
expressions $a^{i}(\mathbf{q})$. Corresponding Riemann invariants $\lambda
^{i}(\mathbf{a})$ are given by (\ref{water})%
\begin{equation}
\lambda ^{i}=q^{i}-\sum_{m=1}^{M}\epsilon _{m}\ln (q^{i}-a^{m}).  \label{rim}
\end{equation}%
Other $N-M$ Riemann invariants $r^{j}(\mathbf{a,b})$ (see (\ref{reda})) are
mark points on the aforementioned Riemann surface%
\begin{equation*}
r^{j}\equiv \lambda |_{q=b^{j}}=b^{j}-\sum_{m=1}^{M}\epsilon _{m}\ln
(b^{j}-a^{m}).
\end{equation*}%
Since $\tilde{\lambda}_{j}$ are arbitrary functions of corresponding Riemann
invariants (i.e. $\tilde{\lambda}_{j}=Q_{j}(r^{j})$), the function $a$ (see (%
\ref{y})) can be found from the system%
\begin{equation}
\partial _{i}\partial _{j}a+\frac{\epsilon _{i}}{a^{i}-a^{j}}\partial
_{j}\delta a-\frac{\epsilon _{j}}{a^{i}-a^{j}}\partial _{i}\delta a=\epsilon
_{i}\epsilon _{j}\sum_{m=1}^{N-M}\frac{Q_{m}^{\prime }}{%
(a^{i}-b^{m})(a^{j}-b^{m})},\text{ \ }i\neq j,  \label{re}
\end{equation}%
where (see (\ref{bio}))%
\begin{equation}
\tilde{\partial}_{j}a=\tilde{\lambda}_{j}=Q_{j}\left(
b^{j}-\sum_{m=1}^{M}\epsilon _{m}\ln (b^{j}-a^{m})\right) .  \label{lya}
\end{equation}%
To avoid a complexity of further computations (an integration of arbitrary
functions $Q_{j}(r^{j})$ in (\ref{lya})), let us restrict our consideration
on a first nontrivial case $Q_{j}(r^{j})=\beta _{j}r^{j}$, where $\beta _{j}$
are arbitrary constants. Then (\ref{lya}) leads to%
\begin{equation}
a=B(\mathbf{a})+\frac{1}{2}\sum_{n=1}^{N-M}\beta
_{n}(b^{n})^{2}+\sum_{m=1}^{M}\sum_{n=1}^{N-M}\epsilon _{m}\beta
_{n}(a^{m}-b^{n})[\ln (a^{m}-b^{n})-1],  \label{d}
\end{equation}%
where the function $B(\mathbf{a})$ satisfies (see (\ref{re}))%
\begin{equation*}
\partial _{i}\partial _{j}B+\frac{\epsilon _{i}}{a^{i}-a^{j}}\partial
_{j}\delta B-\frac{\epsilon _{j}}{a^{i}-a^{j}}\partial _{i}\delta B=0,\text{
\ }i\neq j.
\end{equation*}%
This linear system possesses a general solution parameterized by $N-M$
arbitrary functions of a single variable. Its first nontrivial solution is
given by%
\begin{equation}
B=\frac{\delta }{2}\sum_{m=1}^{M}\epsilon
_{m}(a^{m})^{2}+\sum_{m=1}^{M}\gamma _{m}a^{m},  \label{q}
\end{equation}%
where $\delta $ and $\gamma _{k}$ are new arbitrary constants.

Thus, the first nontrivial solution of extended Gibbons--Tsarev system (\ref%
{13})--(\ref{15}), (\ref{qwa}), (\ref{comp}) given by (\ref{anc}), (\ref{d})
and (\ref{q}) leads to the first nontrivial $N$ component hydrodynamic
reduction (\ref{c}), (\ref{aj}) of the Manakov--Santini system. In such a
case, the generating function of conservation law densities can be found
explicitly%
\begin{equation*}
\ln p=-\frac{\delta }{2}q^{2}-\delta \sum_{m=1}^{M}\epsilon
_{m}(a^{m}-q)-\sum_{m=1}^{M}\gamma _{m}\ln (a^{m}-q)-q\sum_{n=1}^{N-M}\beta
_{n}\ln (b^{n}-q)-\sum_{n=1}^{N-M}\beta _{n}(b^{n}-q)
\end{equation*}%
\begin{equation}
+\frac{1}{4}\sum_{m=1}^{M}\sum_{n=1}^{N-M}\beta _{n}\epsilon _{m}[\ln
^{2}(b^{n}-q)+2\ln (a^{m}-q)\cdot \ln (b^{n}-q)-\ln ^{2}(a^{m}-q)]  \label{p}
\end{equation}%
\begin{equation*}
-\ln \left( 1+\sum_{m=1}^{M}\frac{\epsilon _{m}}{a^{m}-q}\right) +\frac{1}{2}%
\sum_{m=1}^{M}\sum_{n=1}^{N-M}\beta _{n}\epsilon _{m}\left( \text{Li}_{2}%
\frac{a^{m}-q}{b^{n}-q}-\text{Li}_{2}\frac{b^{n}-q}{a^{m}-q}\right) ,
\end{equation*}%
where the bi-logarithm%
\begin{equation*}
\text{Li}_{2}z=\sum_{k=1}^{\infty }\frac{z^{k}}{k^{2}}.
\end{equation*}%
A corresponding expression for the function $F$ is given by%
\begin{equation*}
F=\frac{\delta }{2}q^{2}+\delta \sum_{m=1}^{M}\epsilon
_{m}(a^{m}-q)+\sum_{m=1}^{M}\gamma _{m}\ln (a^{m}-q)+q\sum_{n=1}^{N-M}\beta
_{n}\ln (b^{n}-q)+\sum_{n=1}^{N-M}\beta _{n}(b^{n}-q)
\end{equation*}%
\begin{equation}
+\frac{1}{4}\sum_{m=1}^{M}\sum_{n=1}^{N-M}\beta _{n}\epsilon _{m}[\ln
^{2}(a^{m}-q)-2\ln (a^{m}-q)\cdot \ln (b^{n}-q)-\ln ^{2}(b^{n}-q)]
\label{ef}
\end{equation}%
\begin{equation*}
+\frac{1}{2}\sum_{m=1}^{M}\sum_{n=1}^{N-M}\beta _{n}\epsilon _{m}\left( 
\text{Li}_{2}\frac{b^{n}-q}{a^{m}-q}-\text{Li}_{2}\frac{a^{m}-q}{b^{n}-q}%
\right) .
\end{equation*}

\section{Generalized hodograph method}

In comparison with integrable hydrodynamic chains their semi-Hamiltonian
hydrodynamic reductions possess $N$ infinite series of conservation laws and
commuting flows (cf. (\ref{e}) and (\ref{vert})). Let us remind, that
hydrodynamic type system (\ref{c}), (\ref{aj}) is associated with the
Riemann surface $\lambda =\Lambda (\mathbf{a};q)$, whose $M$ branch points $%
\lambda ^{i}=\Lambda (\mathbf{a};q^{i})$ determined by $M$ solutions $q^{i}(%
\mathbf{a})$ of the algebraic equation $\partial \Lambda (\mathbf{a}%
;q)/\partial q=0$ are first $M$ Riemann invariants; all other $N-M$ Riemann
invariants $r^{j}=\Lambda (\mathbf{\lambda };b^{j})$ are nothing else but
just mark points on this Riemann surface.

Since a dependence $p(q,\mathbf{a,b})$ can be inverted to $q(p,\mathbf{a,b})$
for any hydrodynamic reduction (\ref{c}), (\ref{aj}), generating function of
conservation laws (\ref{geni})%
\begin{equation*}
\partial _{t}p=\partial _{x}[(q(p)-a)p]
\end{equation*}%
leads to $N$ infinite series of conservation laws (see \cite{algebra}) by
virtue of formal expansions%
\begin{eqnarray}
p^{(i)} &=&p_{0}^{i}+p_{(1)}^{i}\lambda ^{(i)}+p_{(2)}^{i}(\lambda
^{(i)})^{2}+p_{(3)}^{i}(\lambda ^{(i)})^{3}+...,\text{ \ }i=1,2,...,M, 
\notag \\
&&  \label{i} \\
\tilde{p}^{(j)} &=&\tilde{p}_{0}^{j}+\tilde{p}_{(1)}^{j}\tilde{\lambda}%
^{(j)}+\tilde{p}_{(2)}^{j}(\tilde{\lambda}^{(j)})^{2}+\tilde{p}_{(3)}^{j}(%
\tilde{\lambda}^{(j)})^{3}+...,\text{ \ }j=1,2,...,N-M,  \notag
\end{eqnarray}%
where $\lambda ^{(i)}$ and $\tilde{\lambda}^{(j)}$ are local parameters at
the vicinities of $p_{0}^{i}$ and $\tilde{p}_{0}^{j}$, respectively.
However, in such a case, dependencies $q(p_{0}^{i})$ as well as $q(\tilde{p}%
_{0}^{j})$ are highly complicated, because the dependence $p(q)$ is much
more simpler than $q(p)$ in all known examples (see the previous Section).
Thus, instead (\ref{i}), we utilize another formal expansions given by%
\begin{eqnarray}
q^{(i)} &=&a^{i}+a_{(1)}^{i}\lambda ^{(i)}+a_{(2)}^{i}(\lambda
^{(i)})^{2}+a_{(3)}^{i}(\lambda ^{(i)})^{3}+...,\text{ \ }i=1,2,...,M, 
\notag \\
&&  \label{h} \\
\tilde{q}^{(j)} &=&b^{j}+b_{(1)}^{j}\tilde{\lambda}^{(j)}+b_{(2)}^{j}(\tilde{%
\lambda}^{(j)})^{2}+b_{(3)}^{j}(\tilde{\lambda}^{(j)})^{3}+...,\text{ \ }%
j=1,2,...,N-M,  \notag
\end{eqnarray}%
where $\lambda ^{(i)}$ and $\tilde{\lambda}^{(j)}$ are local parameters at
the vicinities of $a^{i}$ and $b^{j}$, respectively. Since, the dependence $%
p(q)$ possesses $M$ singularities in points $a^{i}$ and $N-M$ singularities
in points $b^{j}$, formal expansions (\ref{i}) cannot exist in a general
case. Nevertheless, in some special cases, first $N$ conservation laws can
be found in the form%
\begin{equation}
\partial _{t}p_{0}^{i}=\partial _{x}[(a^{i}-a)p_{0}^{i}],\text{ \ \ \ \ }%
\partial _{t}\tilde{p}_{0}^{j}=\partial _{x}[(b^{j}-a)\tilde{p}_{0}^{j}].
\label{s}
\end{equation}%
Below we restrict our consideration on a simplest sub-case (\ref{simple}),
which is compatible with \textbf{any} reduction of $\lambda (p)$ (see the
previous Section). In this case, $N$ infinite series of conservation laws
can be found utilizing (\ref{h}). It means that first $N$ conservation laws
are determined by (\ref{s}).

Without loss of generality, let us choose the equation for Riemann surface (%
\ref{water}). In such a case (see (\ref{simple}) and (\ref{anc})), three
dimensional four component hydrodynamic type system (\ref{6}) reduces to the
pair of commuting $N$ component hydrodynamic type systems (see (\ref{sem})
and (\ref{c}), (\ref{aj}))%
\begin{eqnarray*}
a_{t}^{i} &=&(a^{i}-a)a_{x}^{i}+u_{x},\text{ \ \ \ \ \ \ \ }%
a_{y}^{i}=((a^{i})^{2}-aa^{i}-c)a_{x}^{i}+a^{i}u_{x}+u_{t},\text{ \ }%
i=1,2,...,M, \\
&& \\
b_{t}^{j} &=&(b^{j}-a)b_{x}^{j}+u_{x},\text{ \ \ \ \ \ \ \ }%
b_{y}^{j}=((b^{j})^{2}-ab^{j}-c)b_{x}^{j}+b^{j}u_{x}+u_{t},\text{ \ }%
j=1,2,...,N-M,
\end{eqnarray*}%
where ($\epsilon =\Sigma \epsilon _{m},\beta =\Sigma \beta _{n},\gamma
=\Sigma \gamma _{m}$)%
\begin{equation*}
b=\frac{1}{2}\sum_{m=1}^{M}\epsilon _{m}(a^{m})^{2}+\epsilon u,\text{ \ \ }c=%
\frac{1}{2}\sum_{m=1}^{M}\gamma _{m}(a^{m})^{2}+\frac{1}{2}%
\sum_{n=1}^{N-M}\beta _{n}(b^{n})^{2}-\frac{1}{2}a^{2}+(\gamma +\beta -1)u.
\end{equation*}

Then all coefficients $p_{(k)}^{i}$ and $\tilde{p}_{(n)}^{j}$ can be found
iteratively, while coefficients $a_{(k)}^{i}(\mathbf{a,b})$ and $b_{(n)}^{j}(%
\mathbf{a,b})$ are determined by formal expansions of equation $\lambda
=\Lambda (\mathbf{a};q)$ of a Riemann surface. A substitution of these
expansions $q^{(i)}$ and $\tilde{q}^{(j)}$ together with other four
expansions%
\begin{eqnarray*}
\partial _{\tau (\zeta )} &=&\partial _{t^{0,i}}+\lambda ^{(i)}\partial
_{t^{1,i}}+(\lambda ^{(i)})^{2}\partial _{t^{2,i}}+(\lambda
^{(i)})^{3}\partial _{t^{3,i}}+..., \\
&& \\
s^{(i)} &=&s_{(0)}^{i}+s_{(1)}^{i}\lambda ^{(i)}+s_{(2)}^{i}(\lambda
^{(i)})^{2}+s_{(3)}^{i}(\lambda ^{(i)})^{3}+..., \\
&& \\
\tilde{\partial}_{\tau (\zeta )} &=&\partial _{y^{0,j}}+\tilde{\lambda}%
^{(j)}\partial _{y^{1,j}}+(\tilde{\lambda}^{(j)})^{2}\partial _{y^{2,j}}+(%
\tilde{\lambda}^{(j)})^{3}\partial _{y^{3,j}}+..., \\
&& \\
\tilde{s}^{(j)} &=&\tilde{s}_{(0)}^{j}+\tilde{s}_{(1)}^{j}\tilde{\lambda}%
^{(j)}+\tilde{s}_{(2)}^{j}(\tilde{\lambda}^{(j)})^{2}+\tilde{s}_{(3)}^{j}(%
\tilde{\lambda}^{(j)})^{3}+...
\end{eqnarray*}%
to (\ref{e}) leads to $N$ infinite series of commuting flows. Then the
generalized hodograph method admits to construct infinitely many particular
solutions in an implicit form (see \cite{Tsar}).

\section{Conclusion}

Manakov--Santini system (\ref{6}) is the first example in the theory of
three dimensional hydrodynamic type systems, which is naturally equipped by
the two component pseudo-differentials $q,f$, determined by systems in
partial derivatives of the first order%
\begin{eqnarray*}
\left( 
\begin{array}{c}
q \\ 
f%
\end{array}%
\right) _{t} &=&(q-a)\left( 
\begin{array}{c}
q \\ 
f%
\end{array}%
\right) _{x}+\left( 
\begin{array}{c}
u \\ 
a%
\end{array}%
\right) _{x}, \\
&& \\
\left( 
\begin{array}{c}
q \\ 
f%
\end{array}%
\right) _{y} &=&(q^{2}-aq-c)\left( 
\begin{array}{c}
q \\ 
f%
\end{array}%
\right) _{x}+q\left( 
\begin{array}{c}
u \\ 
a%
\end{array}%
\right) _{x}+\left( 
\begin{array}{c}
u \\ 
a%
\end{array}%
\right) _{t}.
\end{eqnarray*}%
We believe that more complicated vector pseudo-differentials%
\begin{equation*}
\vec{f}_{t}=\hat{G}(f,\mathbf{u})\vec{f}_{x}+\hat{F}(f,\mathbf{u})\vec{u}_{x}
\end{equation*}%
are associated with vector hydrodynamic chains (\ref{vek}).

\section*{Acknowledgement}

Authors thank Leonid Bogdanov, Eugeni Ferapontov, Igor Krichever, Oleg
Morozov, Alexander Odesskii and Sergey Tsarev for their stimulating and
clarifying discussions.

MVP is grateful to the Institute of Mathematics in Taipei (Taiwan) where
some part of this work has been done. This research was particularly
supported by the Russian--Taiwanese grant 95WFE0300007 (RFBR grant
06-01-89507-HHC) and by the grant of Presidium of RAS \textquotedblleft
Fundamental Problems of Nonlinear Dynamics\textquotedblright .

JHC was particularly supported by the National Science Council of Taiwan
under the grant No. NSC 96-2115-M-606-001-MY2; YTC was particularly
supported by the National Science Council of Taiwan under the grant No. NSC
97-2811-M-606-001.

\end{document}